\input harvmac
\input epsf
\input amssym 

\overfullrule=0pt
%
%
\def\nextline{\hfil\break}

\font\manual=manfnt \def\dbend{\lower3.5pt\hbox{\manual\char127}}

\def\ie{{\it i.e.}}
\def\eg{{\it e.g.}}
\def\cf{{\it c.f.}}
\def\etal{{\it et.al.}}

\def\sst{\scriptscriptstyle}

\def\frac#1#2{{#1\over#2}}
\def\coeff#1#2{{\textstyle{#1\over #2}}}
\def\half{\frac12}
\def\hf{{\textstyle\half}}
\def\d{\partial}
\def\p{\partial}

\def\inbar{\,\vrule height1.5ex width.4pt depth0pt}
\def\IR{\relax{\rm I\kern-.18em R}}
\def\IC{\relax\hbox{$\inbar\kern-.3em{\rm C}$}}
\def\IQ{\relax\hbox{$\inbar\kern-.3em{\rm Q}$}}
\def\IH{\relax{\rm I\kern-.18em H}}
\def\IN{\relax{\rm I\kern-.18em N}}
\def\IP{\relax{\rm I\kern-.18em P}}
\font\cmss=cmss10
\font\cmsss=cmss10 at 7pt
\def\IZ{\relax\ifmmode\mathchoice
{\hbox{\cmss Z\kern-.4em Z}}{\hbox{\cmss Z\kern-.4em Z}}
{\lower.9pt\hbox{\cmsss Z\kern-.4em Z}}
{\lower1.2pt\hbox{\cmsss Z\kern-.4em Z}}\else{\cmss Z\kern-.4em
Z}\fi}                                                           
\def\Z{{\IZ}}
\def\One{{1\hskip -3pt {\rm l}}}

\catcode`\@=11
\def\slash#1{\mathord{\mathpalette\c@ncel{#1}}}
\def\underrel#1\over#2{\mathrel{\mathop{\kern\z@#1}\limits_{#2}}}

\catcode`\@=12
%
%
\def\sdtimes{\rtimes}
\def\ket#1{|#1\rangle}
\def\bra#1{\langle#1|}
\def\vev#1{\langle#1\rangle}
\def\det{{\rm det}}

\def\mod{{\rm mod}}

\def\det{{\rm det}}

\def\exp{{\rm exp}}
\def\ker{{\mathop{\rm ker}}}

\def\dim{{\mathop{\rm dim}}}

\def\Aut{{\rm Aut}}
\def\Hom{{\rm Hom}}
%
%
\def\AA{{\cal A}} 
\def\BB{{\cal B}} 
\def\CC{{\cal C}} 

\def\GG{{\cal G}} 
\def\HH{{\cal H}} 
\def\II{{\cal I}} 
 
\def\KK{{\cal K}} 
 
\def\MM{{\cal M}} 
\def\NN{{\cal N}} 
\def\OO{{\cal O}} 
\def\PP{{\cal P}} 
 
\def\RR{{\cal R}} 
\def\SS{{\cal S}} 
\def\TT{{\cal T}} 
\def\UU{{\cal U}} 
\def\VV{{\cal V}} 
\def\WW{{\cal W}} 
\def\XX{{\cal X}} 
\def\YY{{\cal Y}}

\def\vareps{\varepsilon}
%
%
\def\unlockat{\catcode`\@=11}
\def\lockat{\catcode`\@=12}
\unlockat
\def\newsec#1{\global\advance\secno by1\message{(\the\secno. #1)}
\global\subsecno=0\global\subsubsecno=0\eqnres@t\noindent
{\bf\the\secno. #1}
\writetoca{{\secsym} {#1}}\par\nobreak\medskip\nobreak}
\global\newcount\subsecno \global\subsecno=0
\def\subsec#1{\global\advance\subsecno
by1\message{(\secsym\the\subsecno. #1)}
\ifnum\lastpenalty>9000\else\bigbreak\fi\global\subsubsecno=0
\noindent{\it\secsym\the\subsecno. #1}
\writetoca{\string\quad {\secsym\the\subsecno.} {#1}}
\par\nobreak\medskip\nobreak}
\global\newcount\subsubsecno \global\subsubsecno=0
\def\subsubsec#1{\global\advance\subsubsecno by1
\message{(\secsym\the\subsecno.\the\subsubsecno. #1)}
\ifnum\lastpenalty>9000\else\bigbreak\fi
\noindent\quad{\secsym\the\subsecno.\the\subsubsecno.}{#1}
\writetoca{\string\qquad{\secsym\the\subsecno.\the\subsubsecno.}{#1}}
\par\nobreak\medskip\nobreak}
\def\subsubseclab#1{\DefWarn#1\xdef
#1{\noexpand\hyperref{}{subsubsection}%
{\secsym\the\subsecno.\the\subsubsecno}%
{\secsym\the\subsecno.\the\subsubsecno}}%
\writedef{#1\leftbracket#1}\wrlabeL{#1=#1}}
\lockat
%
%
\def\G{G}	
\def\P{{\bf P}}	
\def\T{{\bf T}}	
	
\def\IF{\relax{\rm I\kern-.18em F}}

\def\dlat{\relax\ifmmode\mathchoice
{\hbox{\cmss D}}{\hbox{\cmss D}}
{\lower.9pt\hbox{\cmsss D}}
{\lower1.2pt\hbox{\cmsss D}}\else{\cmss D}\fi}
\def\pia{\pi_\AA^{ }}
\def\pig{\pi_\G^{ }}
\def\pirho{\PP_\rho^{ }}
\def\pirhop{\PP_{\rho'}^{ }}

\def\matn{{\sl Mat}_N}

\def\ap{{\alpha'}}

\def\fix{{\rm fix}}

\def\str{{\rm str}}

\def\ncint{\relax{\int \kern-1.06em -}}
\def\aa{{\bf a}}
\def\bb{{\bf b}}
\def\cc{{\bf c}}
\def\dd{{\bf d}}
\def\ff{{\bf f}}
\def\wl{\tilde{\dlat}}
%
%
%
\lref\gp{E.~Gimon and J.~Polchinski, 
``Consistency Conditions for Orientifolds and D-Manifolds,''
Phys.\ Rev.\  {\bf D54}, 1667 (1996)
hep-th/9601038.}
\lref\dm{M.~R.~Douglas and G.~Moore,
``D-branes, Quivers, and ALE Instantons,'' hep-th/9603167.}
\lref\gijo{E.~G.~Gimon and C.~V.~Johnson,
``K3 Orientifolds,''
Nucl.\ Phys.\ {\bf B477}, 715 (1996)
hep-th/9604129.}
\lref\poltens{J.~Polchinski,
``Tensors from K3 orientifolds,''
Phys.\ Rev.\ D {\bf 55}, 6423 (1997)
hep-th/9606165.}
\lref\jomy{C.~V.~Johnson and R.~C.~Myers,
``Aspects of type IIB theory on ALE spaces,''
Phys.\ Rev.\  {\bf D55}, 6382 (1997) hep-th/9610140.}
\lref\bcd{D.~Berenstein, R.~Corrado and J.~Distler,
``Aspects of ALE matrix models and twisted matrix strings,''
Phys.\ Rev.\  {\bf D58}, 026005 (1998)
hep-th/9712049.}
\lref\mrdtors{M.~R.~Douglas,
``D-branes and discrete torsion,''
hep-th/9807235;
M.~R.~Douglas and B.~Fiol,
``D-branes and discrete torsion. II,''
hep-th/9903031.}
\lref\ddg{D.-E. Diaconescu, M. R. Douglas, J. Gomis,
``Fractional Branes and Wrapped Branes'', JHEP {\bf 9802}:013 (1998);
hep-th/9712230.}
\lref\mrdtrieste{M.R.~Douglas,
``Two lectures on D-geometry and noncommutative geometry'',
lectures at the 1998 ICTP Spring School, hep-th/9901146.}
\lref\grossnekrasov{D.~J.~Gross and N.~A.~Nekrasov,
``Solitons in noncommutative gauge theory,''
hep-th/0010090; ``Dynamics of Strings in Noncommutative Gauge Theory,''
JHEP 0010 (2000) 021, hep-th/0007204, ; 
`` Monopoles and Strings in Noncommutative Gauge Theory,''
  JHEP 0007 (2000) 034,hep-th/0005204}
\lref\nekrasov{N.~A.~Nekrasov,
``Trieste lectures on solitons in noncommutative gauge theories,''
hep-th/0011095.}
\lref\koscrev{A.~Konechny and A.~Schwarz,
``Introduction to M(atrix) theory and noncommutative geometry,''
hep-th/0012145.}
\lref\gms{R. Gopakumar, S. Minwalla, and A. Strominger, ``Noncommutative
Solitons,'' JHEP {\bf 0005}:020,2000; hep-th/0003160.}         
\lref\hklm{J. Harvey, P. Kraus, F. Larsen, and E. Martinec,
``D-branes and Strings as Non-commutative Solitons,''
JHEP{\bf 0007}:042, 2000; hep-th/0005031.}                 
\lref\abs{Atiyah, M. F., Bott, R., Shapiro, A.  ``Clifford modules,''
Topology 3 1964 suppl. 1, 3--38. }
\lref\wittenk{E. Witten, ``$D$-Branes And $K$-Theory,''
JHEP {\bf 9812}:019, 1998; hep-th/9810188.}      
\lref\wittach{E.~Witten,
``Noncommutative tachyons and string field theory,''
hep-th/0006071.}
\lref\wittensft{E.~Witten,
``Noncommutative Geometry And String Field Theory,''
Nucl.\ Phys.\ {\bf B268} (1986) 253.}
\lref\bankspeskin{T.~Banks and M.~E.~Peskin,
``Gauge Invariance Of String Fields,''
Nucl.\ Phys.\ {\bf B264}, 513 (1986).}
\lref\lpp{A.~LeClair, M.~E.~Peskin and C.~R.~Preitschopf,
``String Field Theory On The Conformal Plane. 1. Kinematical Principles,''
Nucl.\ Phys.\ {\bf B317}, 411 (1989);
``String Field Theory On The Conformal Plane. 2. Generalized Gluing,''
Nucl.\ Phys.\ {\bf B317}, 464 (1989).}
\lref\sw{N. Seiberg and E. Witten, ``String Theory and Noncommutative
Geometry,'' JHEP {\bf 9909}:032,1999; hep-th/9908142.}             
\lref\wittenstrings{E. Witten, ``Overview of K-theory applied to
strings,'' hep-th/0007175.}                 
\lref\horava{P. Horava, ``Type II D-Branes, K-Theory, and Matrix
Theory,'' Adv. Theor. Math. Phys. {\bf 2} (1999) 1373; hep-th/9812135.}    
\lref\hm{J. Harvey and G. Moore, ``Noncommutative Tachyons and K-Theory,''
hep-th/0009030.}
\lref\matsuo{Y.~Matsuo,
``Topological charges of noncommutative soliton,''
hep-th/0009002.}
\lref\ft{E. S. Fradkin and A.A. Tseytlin, ``Nonlinear 
Electrodynamics From Quantized Strings,'' 
Phys. Lett. {\bf 163B} (1985) 123.}
\lref\acny{A. Abouelsaood, C. G.Callan, C. R. Nappi and S. A. Yost,
``Open Strings in Background Gauge Fields,'' Nucl. Phys.
{\bf B280} (1987) 599.}                                              
\lref\cds{A. Connes, M. R. Douglas and A. Schwarz, ``Noncommutative Geometry
and Matrix Theory: Compactification on Tori,'' JHEP {\bf 9802}:003 (1998);
hep-th/9711162.}
\lref\doughull{M.R.~Douglas and C.~Hull,
``D-branes and the Noncommutative Torus'',
JHEP {\bf 9802} (1998) 008; hep-th/9711165.}
\lref\schmo{A.~Schwarz,
``Morita equivalence and duality,''
Nucl.\ Phys.\ {\bf B534}, 720 (1998);
hep-th/9805034.}
\lref\mz{B.~Morariu and B.~Zumino,
``Super Yang-Mills on the noncommutative torus,''
hep-th/9807198; 
D.~Brace, B.~Morariu, B.~Zumino,
``Dualities of the Matrix Model from T-Duality of the Type II String''
Nucl.Phys. {\bf B545} (1999) 192, hep-th/9810099;
``T-Duality and Ramond-Ramond Backgrounds in the Matrix Model
Nucl.Phys. {\bf B549} (1999) 181, hep-th/9811213.}
\lref\schomerus{V. Schomerus, ``D-branes and Deformation Quantization,''
JHEP {\bf 9906}:030 (1999); hep-th/9903205.}                                    
\lref\ks{V. A. Kostelecky and S. Samuel, ``On a Nonperturbative Vacuum
for the Open Bosonic String,'' Nucl. Phys. {\bf B336} (1990) 263.}
\lref\swedes{U. Lindstrom, M. Rocek, R. von Unge,
``Non-commutative Soliton Scattering'', hep-th/0008108.}
\lref\senissues{See A.~Sen,
``Some issues in non-commutative tachyon condensation,''
hep-th/0009038, and references therein.}
\lref\sendbi{A.~Sen,
``Supersymmetric world-volume action for non-BPS D-branes,''
JHEP{\bf 9910}, 008 (1999)
hep-th/9909062.}
\lref\ripo{M.A.~Rieffel,
``$C^*$ algebras associated with irrational rotations'',
Pac.\ J.\ Math. {\bf 93} (1981), 415.}
\lref\rican{M.A.~Rieffel,
``The cancellation theorem for projective modules over 
irrational rotation $C^{*}$-algebras'',
Proc. London Math. Soc. (3)
{\bf 47} (1983) 285-302.}
\lref\schwsusy{A.~Astashkevich and A.~Schwarz,
``Projective modules over non-commutative tori: 
classification of modules with constant curvature connection,''
math.qa/9904139.}
\lref\connescrasp{A.~Connes,
``$C^*$ algebras and differential geometry'',
Compt. Rend. Acad. Sci. Ser. A {\bf 290} (1980) 599;
translated from original French journal article in 
hep-th/0101093.}
\lref\connes{A. Connes, {\it Noncommutative Geometry}, 
Academic Press (1994).}
\lref\varilly{J.~C.~V\'arilly, 
``An introduction to noncommutative geometry'',
physics/9709045.}
\lref\brodzki{J.~Brodzki,
``An Introduction to K-theory and Cyclic Cohomology'',
funct-an/9606001.}
\lref\landi{G.~Landi,
``An Introduction to Noncommutative Spaces and their Geometry'',
Lecture Notes in Physics: Monographs, m51 
(Springer-Verlag, Berlin Heidelberg, 1997) ISBN 3-540-63509-2;
hep-th/9701078.}
\lref\koscorb{A.~Konechny and A.~Schwarz,
``Compactification of M(atrix) theory on noncommutative toroidal orbifolds,''
Nucl.\ Phys.\  {\bf B591}, 667 (2000),
hep-th/9912185;
``Moduli spaces of maximally supersymmetric solutions 
on noncommutative  tori and noncommutative orbifolds,''
JHEP {\bf 0009}, 005 (2000), hep-th/0005174.}
\lref\hkl{J.~Harvey, P.~Kraus, and F.~Larsen, 
``Exact noncommutative solitons'', hep-th/0010060.}
\lref\rieffel{Rieffel, ``Projective modules over higher-dimensional
noncommutative tori,''
Can. J. Math. {\bf 40} (1988) 257.}
\lref\elliott{G.A.~Elliott,
``On the K-theory of the $C^*$-algebra generated by a projective
representation of a torsion-free discrete group,''
in {\it Operator algebras and group representations, vol. I}, 
G. Arsene \etal, eds.; Pitman Publishing (1984).}
\lref\myers{R.~C.~Myers, ``Dielectric-branes,''
JHEP {\bf 9912}, 022 (1999); hep-th/9910053.}
\lref\periwal{V.~Periwal, ``D-brane charges and K-homology,''
JHEP {\bf 0007}, 041 (2000), hep-th/0006223;
``Deformation quantization as the origin of D-brane 
non-Abelian degrees  of freedom,''
JHEP {\bf 0008}, 021 (2000), hep-th/0008046.}
\lref\kosc{A.~Konechny and A.~Schwarz,
``Supersymmetry algebra and BPS states of super Yang-Mills 
theories on  noncommutative tori,''
Phys.\ Lett.\  {\bf B453}, 23 (1999)
hep-th/9901077.}
\lref\pioline{B.~Pioline and A.~Schwarz, 
``Morita equivalence and T-duality (or B versus Theta),''
JHEP {\bf 9908}, 021 (1999); hep-th/9908019.}
\lref\seiberg{N.~Seiberg,
``A note on background independence in noncommutative gauge theories,  
matrix model and tachyon condensation,''
JHEP {\bf 0009}, 003 (2000);
hep-th/0008013.}
\lref\ryang{S.~Ryang,
``Open string and Morita equivalence of the Dirac-Born-Infeld 
action with  modulus Phi,''
hep-th/0003204.}
\lref\hv{C.~Hofman and E.~Verlinde,
``U-duality of Born-Infeld on the noncommutative two-torus,''
JHEP {\bf 9812}, 010 (1998),
hep-th/9810116;
``Gauge bundles and Born-Infeld on the noncommutative torus,''
Nucl.\ Phys.\  {\bf B547}, 157 (1999),
hep-th/9810219.}
\lref\wittenwzw{E.~Witten, 
``Quantum Field Theory, Grassmannians, And Algebraic Curves,''
Commun.\ Math.\ Phys.\  {\bf 113} (1988) 529;
``Quantum Field Theory And The Jones Polynomial,''
Commun.\ Math.\ Phys.\  {\bf 121} (1989) 351.}
\lref\rieffelheis{M.A.~Rieffel, ``On the uniqueness of the
Heisenberg commutation relations'', Duke Math. J. {\bf 39} (1972) 745.}
\lref\ghi{D.~Gross, A.~Hashimoto, and N.~Itzhaki,
``Observables of non-commutative gauge theories'',
hep-th/0008075.}
\lref\bgr{L.G.~Brown, P.~Green, and M.A.~Rieffel,
``Stable isomorphism and strong Morita equivalence of $C^*$-algebras'',
Pac. J. Math. {\bf 71} (1977) 349.}
\lref\dasrey{S.R.~Das and S.-J. Rey,
``Open Wilson lines in noncommutative gauge theory and 
tomography of holographic dual supergravity'',
hep-th/0008042.}
\lref\walters{S.~Walters,
``Projective modules over the non-commutative sphere'',
J. London Math. Soc. {\bf 51} (1995) 589;
``Chern characters of Fourier modules'',
Can. J. Math. {\bf 52} (2000) 633.}
\lref\bratteli{O.~Bratteli,
``Fixedpoint algebras versus crossed product algebras'',
Proc. Symp. Pure Math. {\bf 38} part 1 (1982) 357-359.}
\lref\boca{F.P.~Boca,
``Projections in rotation algebras and theta functions'',
Comm. Math. Phys. {\bf 202} (1999) 325.}
\lref\bellissard{J.~Bellissard, A.~van~Elst, and H.~Schulz-Baldes,
``The noncommutative geometry of the quantum Hall effect'',
J. Math. Phys. {\bf 35} (1994) 5373;
cond-mat/9411052.}
\lref\ass{J.E.~Avron, R.~Seiler and B.~Simon,
``Charge deficiency, charge transport and comparison of dimensions,''
Commun.\ Math.\ Phys.\ {\bf 159} (1994) 399;
physics/9803014.
}
\lref\kenwilk{C.~Kennedy and A.~Wilkins,
``Ramond-Ramond couplings on brane-antibrane systems,''
Phys.\ Lett.\ {\bf B464}, 206 (1999)
hep-th/9905195.}
\lref\quillen{D.~Quillen,
``Superconnections and the Chern character'',
Topology {\bf 24} (1985) 89;
``Superconnection character forms and the Cayley transform'',
Topology {\bf 27} (1988) 211.}
\lref\jlo{A.~Jaffe, A.~Lesniewski, and K.~Osterwalder,
``Quantum K-theory, I. The Chern character'',
Comm. Math. Phys. {\bf 118} (1988) 1;
A.~Lesniewski and K.~Osterwalder,
``Superspace formulation of the Chern character 
of a theta summable Fredholm module,''
Comm.\ Math.\ Phys.\ {\bf 168} (1995) 643 
hep-th/9406052.}
\lref\CSrefs{M.~Li,
``Boundary States of D-Branes and Dy-Strings,''
Nucl.\ Phys.\ {\bf B460}, 351 (1996) hep-th/9510161;
M.~R.~Douglas, ``Branes within branes,'' hep-th/9512077;
M.~B.~Green, J.~A.~Harvey and G.~Moore,
``I-brane inflow and anomalous couplings on D-branes,''
Class.\ Quant.\ Grav.\ {\bf 14}, 47 (1997) hep-th/9605033;
Y.~E.~Cheung and Z.~Yin,
``Anomalies, branes, and currents,''
Nucl.\ Phys.\ {\bf B517}, 69 (1998)
hep-th/9710206.}
\lref\minmo{R.~Minasian and G.~Moore,
``K-theory and Ramond-Ramond charge,''
JHEP{\bf 9711}, 002 (1997)
hep-th/9710230.}
\lref\mukhi{K.~Dasgupta, S.~Mukhi and G.~Rajesh,
``Noncommutative tachyons,''
JHEP{\bf 0006}, 022 (2000)
hep-th/0005006.}
\lref\gopakumar{Talk delivered by R. Gopakumar at Strings 2001.\nextline
http://theory.theory.tifr.res.in/strings/Proceedings/gkumar/}
\lref\mukhiCS{S.~Mukhi and N.~V.~Suryanarayana,
``Chern-Simons terms on noncommutative branes,''
JHEP{\bf 0011}, 006 (2000)
hep-th/0009101.}
\lref\uvirmin{S.~Minwalla, M.~Van Raamsdonk and N.~Seiberg,
``Noncommutative perturbative dynamics,''
hep-th/9912072.}
\lref\uvirmat{A.~Matusis, L.~Susskind and N.~Toumbas,
``The IR/UV connection in the non-commutative gauge theories,''
hep-th/0002075.}
\lref\olsenszabo{K. Olsen and R.J. Szabo, 
`` Constructing D-Branes from K-Theory,''
hep-th/9907140}
\lref\lark{P.~Kraus and F.~Larsen,
``Boundary String Field Theory of the DDbar System'',
hep-th/0012198.}
\lref\ttu{T.~Takayanagi, S.~Terashima and T.~Uesugi,
``Brane-Antibrane Action from Boundary String Field Theory'',
hep-th/0012210.}
\lref\aio{M.~Alishahiha, H.~Ita and Y.~Oz,
``Superconnections and the tachyon effective action'',
hep-th/0012222.}
\lref\nsv{K.~S.~Narain, M.~H.~Sarmadi and C.~Vafa,
``Asymmetric Orbifolds,''
Nucl.\ Phys.\ {\bf B288}, 551 (1987).}
\lref\schnabl{M.~Schnabl,
``String field theory at large B-field and noncommutative geometry,''
JHEP{\bf 0011}, 031 (2000)
hep-th/0010034.}
\lref\bars{I.~Bars, H.~Kajiura, Y.~Matsuo, and T.~Takayanagi,
``Tachyon condensation on noncommutative torus'',
hep-th/0010101.}
\lref\morocco{El Mostapha Sahraoui and El Hassan Saidi,
``Solitons on compact and noncompact spaces in large noncommutativity'',
hep-th/0012259.}
\lref\morrocco{El~M.~Sahraoui and El~H.~Saidi,
``Solitons on compact and noncompact spaces in large noncommutativity'',
hep-th/0012259.}
\lref\atiyahsegal{M.F.~Atiyah and G.B.~Segal, ``The index
of elliptic operators II,'' Ann. of Math. {\bf 87} (1968) 531.} 
\lref\atiyah{M.~Atiyah and G.~Segal,
``On equivariant Euler characteristics'',
J. Geom. Phys. {\bf 6} (1989) 671.}
\lref\ginsparg{P.~Ginsparg,
``Applied Conformal Field Theory,''
in {\it Fields, Strings, Critical Phenomena},  
ed. by E.~Brezin and J.~Zinn-Justin; North-Holland (1990).}
\lref\asymorbrefs{I.~Brunner, A.~Rajaraman and M.~Rozali,
``D-branes on asymmetric orbifolds,''
Nucl.\ Phys.\ {\bf B558}, 205 (1999),
hep-th/9905024;
B.~Kors,
``D-brane spectra of nonsupersymmetric, asymmetric orbifolds 
and  nonperturbative contributions to the cosmological constant,''
JHEP{\bf 9911}, 028 (1999),
hep-th/9907007;
R.~Blumenhagen, L.~Gorlich, B.~Kors and D.~Lust,
``Asymmetric orbifolds, noncommutative geometry and type I string vacua,''
Nucl.\ Phys.\ {\bf B582}, 44 (2000),
hep-th/0003024;
M.~Gutperle,
``Non-BPS D-branes and enhanced symmetry in an asymmetric orbifold,''
JHEP{\bf 0008}, 036 (2000),
hep-th/0007126;
J.A.~Harvey, S.~Kachru, G.~Moore, and E.~Silverstein, unpublished.}
\lref\paperii{To appear} 
\lref\blackadar{B. Blackadar, {\it K-Theory for Operator
Algebras}, MSRI Publications 5, Cambridge Univ. Press,
1998.}
\lref\raeburn{I. Raeburn and D.P. Williams, {\it Morita 
Equivalence and Continuous-Trace $C^*$-Algebra}, Amer. Math. 
Soc. 1998} 
\lref\hormar{G.~T.~Horowitz and D.~Marolf,
``A new approach to string cosmology,''
JHEP{\bf 9807}, 014 (1998);
hep-th/9805207.}
\lref\hkms{J.~A.~Harvey, S.~Kachru, G.~Moore and E.~Silverstein,
``Tension is dimension,''
JHEP{\bf 0003}, 001 (2000);
hep-th/9909072.}
\lref\irrax{T.~Banks, M.~Dine and N.~Seiberg,
``Irrational axions as a solution of the strong 
CP problem in an eternal universe,''
Phys.\ Lett.\ {\bf B273}, 105 (1991)
hep-th/9109040.}
\lref\simplealg{S.C.~Power,
``Simplicity of $C^*$-algebras of minimal dynamical systems'', 
J. London Math. Soc. (2) {\bf 18} (1978) 534.}

\Title{\vbox{\baselineskip12pt
\hbox{hep-th/0101199}
\hbox{EFI-2000-55}
\hbox{RUNHETC-2000-58} 
}}%
{\vbox{\centerline{Noncommutative Solitons on Orbifolds}
\medskip
}}
 
\smallskip
\centerline{Emil J. Martinec}
\medskip
\centerline{\it Enrico Fermi Institute and Dept. of Physics}
\centerline{\it University of Chicago}
\centerline{\it 5640 Ellis Ave. Chicago IL, 60637}
\bigskip
\centerline{Gregory Moore}
\medskip
\centerline{\it Department of Physics, Rutgers University}
\centerline{\it Piscataway, New Jersey, 08855-0849}
 
\bigskip                                                  
\bigskip
In the noncommutative field theory of open strings in a $B$-field,
D-branes arise as solitons described as projection operators or
partial isometries in a $C^*$ algebra.  
We discuss how D-branes on orbifolds
fit naturally into this algebraic framework, through the examples
of $\IR^n/G$, $\T^n=\IR^n/\IZ^n$, and $\T^n/G$.  
We also propose a framework
for formulating D-branes on asymmetric orbifolds.

\medskip

\Date{January 26, 2001}        

\newsec{Introduction}

Field theory on a noncommutative space is proving to be a useful
limiting case of string theory \refs{\cds,\sw}, 
which still preserves some interesting aspects of
stringy structure, for instance UV/IR mixing 
\refs{\uvirmin,\uvirmat}
and T-duality \refs{\cds,\schmo,\mz,\kosc,\pioline,\hv,\sw}.
In some respects, the scale set by the noncommutativity 
plays a role similar to the string scale -- as both a 
regulator and a source of nonlocality in the theory.
In this setting, the role of the algebra of functions 
on a space is played by a noncommutative $C^*$ algebra $\AA$.  
The simplest example is the noncommutative plane $\IR^{2d}$.  

The algebraic structure of noncommutative
geometry allows a particularly simple description of
D-branes as noncommutative solitons 
\refs{\gms,\mukhi,\hklm}, and clarifies 
\refs{\wittenstrings,\hm,\matsuo}
the relation between D-branes and K-theory 
\refs{\minmo,\wittenk,\horava,\olsenszabo\wittenstrings}.
As we review in section 2, D-branes of even codimension on $\IR^d$
are solitons in the effective field theory of open strings.
The soliton configurations are described by projection
operators or partial isometries in the $C^*$ algebra $\AA$.
One of the goals of this work is to generalize the construction
(and classification) of D-branes as solitons
to the next simplest string target spaces, namely tori and orbifolds.

The study of D-branes on orbifolds was begun 
in \refs{\gp,\dm,\gijo}.  For noncompact orbifolds $\IR^d/\G$, 
the spectrum of branes and the field theories on them
are determined by the representation theory of $\G$ 
\refs{\dm,\gijo,\bcd,\ddg}.
Placing an image D-brane on each leaf of the covering space $\IR^d$
yields a set of branes transforming in the regular representation
of $\G$.  When taken to the orbifold point,
this brane splits up into a set of {\it fractional branes}
according to the decomposition of the regular representation
into irreducible representations; these fractional branes are the basic
objects in the classification of branes on the orbifold.
These basic branes can again be thought of in terms of a
projection of branes on the covering space $\IR^d$.

The appropriate and universal definition 
of the quotient of an algebra $\AA$ by 
the action of a group $\G$ is actually to consider an 
enlarged algebra known as the {\it crossed product} $\AA\sdtimes\G$
\refs{\dm,\mrdtrieste,\koscorb},
which we define in section 3.  
Multiplication of elements in the crossed
product involves the multiplication 
laws in both $\G$ and $\AA$, as well as the action of $\G$ on $\AA$.
Physical string states belong to 
the $\G$-invariant subalgebra of the crossed product.

It is thus a straightforward consequence of this observation,
combined with the construction of \hklm, that given an
algebraic description of D-brane solitons on a space $\YY$
as operators in an algebra $\AA$, then the description
of D-brane solitons on $\YY/\G$ will arise upon taking the crossed
product $\AA\sdtimes\G$, and constructing appropriate
projectors or partial isometries there.
We illustrate this procedure in sections 3 and 4
by constructing projectors and partial isometries for
the orbifolds $\IR^d/\G$, $\T^d=\IR^d/\Z^d$, and $\T^d/\G$.
Tachyon condensation on the two-dimensional noncommutative torus 
has also been discussed recently in \refs{\schnabl,\bars,\morocco}. 

In a companion paper \paperii, we intend to apply the
tools of noncommutative geometry to relate the classification
of D-brane charges to the K-theory and the 
cohomology of the corresponding noncommutative algebras.
The application of noncommutative differential geometry
\connescrasp\ 
(for reviews, see \refs{\connes,\brodzki,\landi,\varilly})
in the context of D-branes on noncommutative spaces has been
developed in the works \refs{\cds,\schmo,\mz,\kosc,\schwsusy,\koscorb}
and we will build on this work. 
We will  show how the Chern character of noncommutative differential
geometry, as defined by A. Connes 
\refs{\connescrasp,\connes}\  
is related to the Chern-Simons couplings of branes.
We will also  exhibit a set of cyclic cocycles 
for certain crossed product
algebras that measure the topological invariants of 
branes on orbifolds, and are thus suitable for use
in constructing their Chern-Simons couplings.
 
Related work by R. Gopakumar and M. Headrick has been announced in 
\gopakumar.


\newsec{Summary of D-branes as noncommutative solitons}


\subsec{Noncommutative geometry}

Suppose we have a closed string background $\XX\times \IR^{2d}$.
Consider some number 
of $D9$-branes in type IIA string theory,
or $D9$-$\overline{D9}$ pairs in type IIB string theory 
(the extension to the bosonic string is obvious).
The effect of turning on a nondegenerate $B$ field along
$\IR^{2d}$ can be absorbed \refs{\cds,\doughull,\schomerus,\sw} into
a description of the dynamics in terms of noncommutative geometry.
For other expository discussions of the application
of noncommutative geometry in this context,
see \refs{\mrdtrieste,\nekrasov,\koscrev}.

The multiplication $*$ of fields depending on noncommutative $\IR^{2d}$
is induced by the algebra of coordinates
\eqn\noncomm{
[x^{2i-1}, x^{2i}] := x^{2i-1} * x^{2i} - x^{2i} * x^{2i-1} = -i \theta_i \ ,
}
where the $\theta^i$, $i=1,...,d$
are the skew eigenvalues of the parameter $\Theta^{ij}$
appearing in the Moyal product.  
Then we have the Moyal algebra of functions $\AA$,
for which we can map any function $f$ to a corresponding operator
using Weyl ordering
\eqn\weyl{
f \rightarrow U(f) := \int d^{2d}p \tilde f(p) e^{i p\cdot \hat x}
}
where $\tilde f$ is the commutative Fourier transform and $\hat x$
is the quantization of $x$ using the standard Heisenberg representation.
If we restrict to functions of rapid decrease the set of operators
$U(f)$ is a C*-algebra isomorphic to the algebra of
compact operators $\KK$ (for any $d$, $\Theta$) \rieffelheis.

It is also convenient to introduce the Hilbert space $\HH$
on which $\AA$ acts, and its standard number basis $\ket{\vec n}$
for the raising and lowering operators 
$a_i^\dagger=(x^{2i}-ix^{2i-1})/\sqrt{2\theta_i}$,
$a_i=(x^{2i}+ix^{2i-1})/\sqrt{2\theta_i}$.
Translations on $\IR^{2d}$ are generated by the unitary operators
\eqn\transgen{\eqalign{
  \TT(z)&=\exp[z^ia^\dagger_i-{\bar z^i}a_i]\cr
  \TT(z)\TT(w)&=e^{\half(z\bar w-w\bar z)}\TT(z+w) \ ;
}}
in particular one has $\TT(z)a\TT^{-1}(z)= a-z$.
Associated to the translation operators are the coherent states 
\eqn\cohst{
  \ket{z}=\TT(z)\ket{0}\ .
}
In sections 3 and 4, we will generalize the discussion
to other noncommutative manifolds besides $\IR^{2d}$:
noncompact orbifolds, noncommutative tori and toroidal orbifolds.


\subsec{Effective $D$-brane dynamics}

Integrating out massive string modes
results in a low energy effective action that is
a noncommutative field theory of the open string tachyon and gauge field
\refs{\cds,\schomerus,\sw}; on the type IIA $D9$-brane, we have
\eqn\openact{\eqalign{
S = {c\over G_s} \int_\XX  d^{10-2d}x \sqrt{G} 
  {\Tr}\Bigl[\hf&
 f(T^2-1)G^{\mu\nu} D_\mu T D_\nu T  - V(T^2-1) \cr
 &- {\coeff14} h(T^2-1)\bigl( F_{\mu\nu}+\Phi_{\mu\nu}\bigr) 
		\bigl(F^{\mu\nu}+\Phi^{\mu\nu}\bigr)
	+\ldots\Bigr]
}}
where $G_s$ and $G_{\mu\nu}$ are the coupling and metric
felt by open strings, \cf\ \refs{\ft,\acny,\sw};    
${\Tr}$ is the trace of the operator on Hilbert space; and $x^\mu$ runs
over both the commuting coordinate directions $x^a$ and noncommuting
coordinate directions $x^i$ on $\XX$.  
The antisymmetric tensor $\Phi$ incorporates the possibility
of noncommutativity of derivatives \seiberg
\eqn\ncder{
  [\d_i,\d_j]=i\Phi_{ij}\ .
}
The functions $f,h,V$ are unknown but satisfy certain 
crucial properties in accord with the conjectures of 
\refs{\senissues,\sendbi}. 
The Dirac-Born-Infeld extension of \openact\ has been considered
for tachyon dynamics in \refs{\sw,\sendbi}; and for noncommutative
gauge theory in \refs{\sw,\hv,\pioline,\ryang}.
Similarly, on the type IIB $D9$-$\overline{D9}$ system, we have
a pair of gauge fields $A^\pm$ acting on a complex tachyon field $T$ as
\eqn\coactz{\eqalign{
  D_\mu T &= \partial_\mu T + i ( A^+_\mu T - T A^-_\mu) \cr
  \overline{D_\mu T} &= \partial_\mu \bar T -i (
	\bar T A^+_\mu - A^-_\mu \bar T )\ , 
}}                  
the tachyon potential is $V(T, \bar T)=U(\bar T T-1)+U(T\bar T -1)$,
and the gauge kinetic term is suitably generalized \hkl.
Here and henceforth and overline denotes the $C^*$ algebra adjoint. 
It is convenient to introduce the gauge field 
(in complex coordinates) in the noncommutative directions
\eqn\Cdef{
  C_j=\theta_j^{-\half} a_j^\dagger+i A_j\quad,\qquad 
  \bar C_j=\theta_j^{-\half}a_j-i \bar A_j\ ,
}
in terms of which $(F+\Phi)_{2j-1,2j}=[C_j,\bar C_j]$
(note that this implies
$\Phi=-\Theta^{-1}$ for $F$ a compact operator \seiberg).
The global minimum of the action 
\eqn\clvac{
  |T|=1\quad,\qquad C_j=\theta_j^{-\half}a_j^\dagger
}
is to be identified with the closed string vacuum, having
no perturbative open string excitations \senissues. 

In the limit $\ap B_{ij} \to \infty$
(or equivalently, $\Theta^{ij}/\ap \to 0$),
by rescaling the coordinates to remove $\theta_i$ from
the star product one sees that the action
reduces to the nonderivative terms.
It turns out that there are interesting soliton solutions
of the equations of motion.
In the IIA case, one finds solutions
in terms of a projection operator \gms\ 
\eqn\gmssol{
T= \One-\P\quad,\qquad \P^2=\P\ .
}
A rank $k$ projection operator gives $U(k)$ gauge symmetry
on the lower-dimensional unstable $D(9-2d)$ brane.
The dynamical degrees of freedom on this lower-dimensional
brane are operators mapping the kernel of $T$ to itself.

For type IIB, the complex tachyon $T$ must satisfy \wittach\
the defining equation of a {\it partial isometry}
\eqn\parti{
  T\bar T T=T\ .
}
The net brane charge is the index of $T$; we will assume 
for simplicity that $T$ has vanishing cokernel.
The dynamical degrees of freedom on the lower dimensional
brane again arise from operators mapping the kernel of $T$ to itself.

In both IIA and IIB situations, the tension and effective actions
of these soliton solutions turn out to be exactly
those of lower-dimensional D-branes \hklm.  One might be concerned
that this remarkable result, seemingly at leading order in the 
expansion in inverse powers of $\alpha'B$, receives corrections
at each order that spoil the precise agreement.
However, recently it has been shown \hkl\ that the structure of the
effective action \openact\ is such that a suitable gauge
field can be found for which \gmssol\ and \parti\ are {\it exact}.%
\foot{Very similar constructions also appeared in \grossnekrasov.} 
The idea is that, starting from the closed string 
vacuum solution \clvac, one may use partial isometries $U$
to construct new solutions to the equations of motion
\eqn\solgen{
  T\rightarrow UT\bar U=U\bar U\quad,\qquad
  C\rightarrow UC\bar U=U\theta^{-\half}a^\dagger\bar U.
}
These are almost gauge transformations, in that 
$U\bar U=\One-\P_n$, $\bar U U=\One-\P_m$ shows that the transformation
is unitary in the orthogonal complement to a
(finite-dimensional) kernel and cokernel; hence
$U$ is `almost' a unitary transformation.  In this sense, the
transformation is a bit like that which generates a zero-size
instanton, or a vertex operator in WZW/Chern-Simons theory \wittenwzw --
namely, pure gauge outside a small `singular' region 
which carries topological winding (hence quantized action);
noncommutativity even smooths the singularity.  
The symmetry properties
allow one to show that the transformation \solgen\ 
generates {\it exact} D-brane solutions of
the equations of motion for {\it any} value of $\alpha'B$,
with the correct value of the tension \hkl.

BPS $D$-branes arise in the brane-antibrane system
from the noncommutative version \hm\ of the ABS construction
\refs{\abs,\wittenk,\horava}.
Let 
\eqn\clifford{
\gamma_i  = \pmatrix{ 0 & \Gamma_i \cr \bar \Gamma_i & 0 \cr}
}
be a representation of the $2d$-dimensional Clifford algebra.
Then
\eqn\teebar{
  T = \frac{1}{\sqrt{\Gamma_i x^i\bar\Gamma_i x^i}}\;\Gamma_i x^i
}
satisfies the partial isometry equation $T \bar T T = T$,
has no cokernel and is of index $1$. 
The analysis of \hm\ (see also \matsuo) 
shows that this tachyon field carries the K-theory charge
of a BPS $D(9-2d)$ brane.

In the framework of noncommutative geometry, the quantization 
of the point particle at the endpoint of the open string 
gives a Hilbert space $\HH$. Open strings with the 
same boundary conditions are elements of $\HH\otimes\HH^\vee$, 
\ie\ a pair of string
endpoints with opposite orientations
because we  are allowed to ignore the string oscillators 
in the large $\alpha' B_{ij}$ limit. On the other hand 
$\HH\otimes\HH^\vee$ is an algebra, corresponding to the 
joining of string endpoints.%
\foot{Technically, $\HH\otimes\HH^\vee$  is the algebra of 
Hilbert-Schmidt operators on Hilbert space. The norm 
completion gives the $C^*$ algebra of compact operators.} 
Upon tachyon condensation, $\HH$ has an orthogonal decomposition
into $\ker(T)$, representing the endpoints of strings
on lower-dimensional branes, and $\ker(T)^\perp$,
representing string endpoints ``off the branes''.
Elements of the algebra $\AA$ split into \wittach
\eqn\algdecomp{
  \pmatrix{\UU&\VV\cr \widetilde\VV&\WW}\ ,
}
with $\UU\in\Aut\bigl(\ker(T)\bigr)$, $\WW\in\Aut\bigl(\ker(T)^\perp\bigr)$,
$\VV\in\Hom\bigl(\ker(T)^\perp,\ker(T)\bigr)$, and 
$\widetilde\VV\in\Hom\bigl(\ker(T),\ker(T)^\perp\bigr)$.
Now $\WW$ is the image under \weyl\
of a function supported outside the location of the 
branes, where the tachyon has condensed to the closed string vacuum.
We thus associate $\ker(T)^\perp$ with string
endpoint configurations in the closed string vacuum.
There are many operators in $\AA$ of the
form $\VV$, $\widetilde\VV$, $\WW$ 
(typically most of $\AA$, in fact),
that do not act only within the subspace $\ker(T)$ of $\HH$ corresponding
to the lower-dimensional branes.  Following this line of reasoning 
(adopted from \wittach), 
these operators correspond to open strings
with one or both endpoints trying to end in the closed string
vacuum.  It was shown in \hklm\
that the related fields in the effective action have string scale
mass after tachyon condensation, and so should not be considered
in the effective action approximation.%
\foot{In the level truncation
approximation of the full string field theory, \ks\ showed that
there are no on-shell modes of the form $\WW$ in the linearized
approximation.}
Physically, the absence of such fields is due to the different
ways Gauss' law can be satisfied at $T=0$ (on the unstable D-brane) 
and $T=1$ (in the closed string vacuum).  In the presence of a $D$-brane,
a string can end on the submanifold the $D$-brane occupies,
and the flux sourced by the string endpoint is carried 
at little energy cost on the $D$-brane
worldvolume.  In the closed string vacuum, there are no light excitations
capable of supporting this flux (they cost energy
scaling as $g_s^{-1}$), and an open string endpoint must find
another open string to bind to.
We will henceforth make the standard 
assumption that stringy dynamics
eliminates the unwanted open string excitations from the spectrum
\senissues. 

There are clearly many ways to condense open string tachyons 
to obtain $k$ lower-dimensional $D$-branes; one can
start with some number $M$ of $D9$-branes or $D9$-$\overline {D9}$ pairs,
and distribute $k$ tachyonic solitons in various ways
among them.  Although the starting point is not unique,
the dynamics on the lower-dimensional branes (and in the
closed string vacuum) is unique, and 
thus there is a sort of ``universality'' in the flows
in configuration space resulting from tachyon condensation.
Descriptions beginning with any number of unstable
branes must therefore be equivalent.%
\foot{The effective fields describing the $D9$-branes parametrize 
the collective modes of a particular unstable field configuration;
as the tachyon condenses, they lose their preferred status
and merge with the totality of string field excitations, 
\cf\ \senissues.}
This is reflected in the fact that, while the algebras 
$\AA$ and $\AA\otimes \matn(\IC)$
are not isomorphic, they are {\it Morita equivalent} --
their representation theories are isomorphic.
Hence their K-theories are the same, and the classification
of lower-dimensional brane charges is unaffected
by the freedom in the description.  We will often
find it useful to take advantage of this freedom -- to add 
arbitrary Chan-Paton factors without affecting 
the effective dynamics -- in what follows.

As pointed out in \refs{\hklm,\wittach}, the above discussion carries
over more or less unchanged to the full string theory, at least 
formally.   The
algebra of open string fields $\AA_{\rm str}$ factorizes \wittach\
as $\AA_{\rm str}\rightarrow \AA_0\times\AA_1$ in the limit under
consideration, where $\AA_0$ is the algebra of vertex
operators with zero momentum in the noncommutative directions, and $\AA_1$
is the algebra $\AA$ of noncommutative functions
considered above.

\subsec{The moduli space of separated branes}

We would like to understand the description 
of the moduli space in the noncommutative framework.
The fact that the noncommutative framework reproduces
the correct low-energy dynamics guarantees that the moduli
space will be the same as the commutative description;
one can simply give an expectation value to the transverse
scalars $A_i$ to move a regular representation brane away
from the origin.  

Consider for simplicity noncommutative $\IR^2$.
A partial isometry for $N$ branes at the origin is
\eqn\TN{
  T_{\sst(N)} =\frac1{\sqrt{a^N{a^\dagger}^N}}\cdot a^N \qquad\quad .
}
Note that $T_{\sst(N)}$ has kernel spanned by $\ket{i}$, $i=0,...,N-1$,
vanishing cokernel,
and obeys $T_{\sst(N)}\bar T_{\sst(N)}=\One$, 
and therefore $\bar T_{\sst(N)} T_{\sst(N)}=\One-\P_N$
(where $\P_N=\sum_{i=0}^{N-1}\ket{i}\bra{i}$ is the standard
level $N$ projector).  

Moving the branes away from the origin
can be achieved by giving an expectation value
$A_z=\sqrt\theta\sum_i z_i\ket{i}\bra{i}$ to the
Goldstone mode $A_z$.  
A suitable family of partial isometries allows the
construction of exact solutions \solgen\ \hkl.
{From} \transgen-\cohst, one sees that the partial isometry
\eqn\TZ{\eqalign{
  T_{\{z\}} &=\frac1{\sqrt{\KK\bar\KK}}\cdot\KK \cr
  \KK &=\prod_{k=0}^{N-1}(a-z_k)\cr
}}
has as kernel the space spanned by the coherent
states $\ket{z_i}$, $i=0,...,N-1$, vanishing cokernel,
and obeys $ T_{\{z\}}\bar T_{\{z\}}=\One$, 
and also $\bar  T_{\{z\}} T_{\{z\}}=\One-\P_{\{z\}}$ 
(where $\P_{\{z\}}$ the projector onto the linear span of the $\ket{z_i}$).
This then is an appropriate partial isometry to describe the
IIB case; $\bar T_{\{z\}} T_{\{z\}}$ 
is the appropriate projector for IIA branes.
For large separation, the associated gauge field is approximately
\eqn\AZ{
  A_z\sim \sum_{i=1}^N z_i\ket{z_i}\bra{z_i}
}
in the space spanned by the $\ket{z_i}$,
up to exponentially small corrections due to the fact
that the coherent states are not quite orthogonal,
$|\vev{z_j|z_i}|^2=\exp[-|z_i-z_j|^2]$.
Making a small fluctuation expansion about this background,
one sees that the kinetic terms $|[A_z,\delta T]|^2$
and $|[A_z,\delta A_a]|^2$ give the right masses to the 
tachyon and gauge field excitations on the lower-dimensional branes.
Note also that the moduli space of partial isometries \TZ,
and thus the moduli space of $N$ D-branes on $\IR^2$,
is manifestly $(\IR^2)^N/S_N$.

Above two dimensions, the description is more complicated
due to the fact that the operators $\Gamma\cdot(x-x_i)$
do not commute with one another, due to both the
matrix structure and the noncommutativity of the $x$'s.  
Nevertheless, it is straighforward to show that the $N^{\rm th}$ 
power of \teebar\ has $N$ dimensional kernel, and describes $N$
branes at the origin.  Furthermore, for $x_i$ well separated 
relative to the scale of the noncommutativity, the product 
\eqn\generalK{
  T=\prod_{i=1}^N\;\frac{1}{\sqrt{\Gamma\cdot(x-x_i)
	\bar\Gamma\cdot(x-x_i)}}\;\Gamma\cdot(x-x_i)
}
is well approximated by the corresponding expression
for commuting $x$.  The latter has unit winding around
each $x_i$ (in the vicinity of each $x_i$, one has the ABS construction
left- and right-multiplied by matrices with no winding
around $x_i$); correspondingly, one sees that the operator
\generalK\ has a kernel consisting of states whose
expectation value for $x$ is concentrated around the $x_i$.
Thus it is an appropriate partial isometry for well-separated branes.

One puzzling feature of \generalK\ is that the partial isometry 
depends on the choice of ordering of the factors. 
The space of partial isometries is thus $(\IR^{2d})^N$,
and one might wonder whether it is correct to quotient
by the symmetric group.
However, different choices do not affect the low energy 
lagrangians on the lower-dimensional branes.
Using the exact construction of \hkl, equation \solgen, 
one can show that the low-energy effective action
on $N$ coincident branes is the correct 
supersymmetric $U(N)$ gauge theory, and the symmetric
group quotient is part of the gauge symmetry.
This indicates that changes in the ordering of factors
in \generalK\ only affect the elements $\VV,\widetilde\VV,\WW$
in \algdecomp, and are not a matter of concern;
nevertheless, it would be nice to understand better the detailed
structure of the higher dimensional moduli space.


\newsec{Noncompact ($\IR^{2d}/\G$) orbifolds}

The algebraic structure of the noncommutative description
of D-branes allows a natural incorporation of the algebraic
structure of D-branes on orbifolds \refs{\dm,\bcd,\ddg}.  
The latter consists of building representations of 
the orbifold group via an appropriate projection of 
the open strings for a collection of D-branes on the covering space.
For noncompact orbifolds, the covering space is the noncommutative plane,
whose associated algebra of operators (and its Hilbert space
representation) carries a natural action 
of the discrete subgroup of rotations by which we wish to orbifold.
In particular, we can build operators \gmssol, \parti\ 
for a collection of D-branes; we will see that these intertwine naturally
with the orbifold action to provide projectors for an arbitrary
collection of (fractional) D-branes on the orbifold space.

Suppose we have a point group $\G$ which is 
a finite subgroup of $O(\IR^{2d})$.
To define D-branes on the orbifold $\IR^{2d}/\G$ one must find:

\item{a.)}
A homomorphism $\alpha: \G \to \Aut(\AA)$;
\item{b.)}
A representation $\pia$ of $\AA$ on a Hilbert space $\HH$;
\item{c.)}
A unitary representation $\pig$ of the group on Hilbert space $\HH$;

\noindent
such that we have a ``covariant representation'':
\eqn\covrep{
\pig(\gamma) \pia(a) \pig(\gamma)^{-1} =\pia( \alpha_\gamma(a))
}
for all $a\in \AA$.
This is simply a generalization of the construction of \dm.
There, a scalar field like $T(x)$
was considered as an operator on a finite dimensional
Chan-Paton space, and one imposed $\pig(\gamma) T(x) \pig(\gamma)^* =
(\alpha_\gamma T)(x) = T(\gamma^{-1} x)$.  The above 
generalizes to $T(x)$ which live in a general $C^*$
algebra, rather than the algebra of $N\times N$ matrices.
In particular, there is an obvious action on
functions $(\gamma\cdot f)(x) = f(\gamma^{-1} x) $ which
defines the automorphism $\alpha_\gamma$ on the algebra
$\AA$:
\eqn\automorph{
\alpha_{\gamma}(U(f)):= U(\gamma\cdot f)\ ;
}
In particular, the operators $a_i$, $a_i^\dagger$ transform
as the coordinates of $\IR^{2d}$ under the action of $\G$.
Mathematically, a covariant representation as above
defines a $C^*$ representation of the
{\it crossed product} algebra $\AA\sdtimes \G$. 

Discrete torsion is determined
by an element of group cohomology; 
as suggested in \dm\ and investigated in detail in \mrdtors,
it can be incorporated if we 
simply replace (a) by a projective unitary representation
determined by a two-cocycle in $H^2(\G,U(1))$.
The generalization of (a)-(c) is called a {\it twisted}
cross product.  The multiplication rule is:
\eqn\twistcross{
\biggl( \sum_\gamma  a_\gamma \gamma\biggr)
\cdot
\biggl( \sum_\gamma  b_\gamma \gamma\biggr) =
\sum_{\gamma,\gamma'} a_\gamma \alpha_{\gamma}(b_{\gamma'})
\sigma(\gamma, \gamma') \gamma\gamma'
}
where $\sigma$ is the $U(1)$-valued group cocycle.

Note that physical open strings are invariant under the action
of $\G$; for example, in an annulus diagram the sum over 
twisted boundary conditions in the closed string channel
amounts to a projection onto $\G$-invariant states in the
open string channel.  In the language of noncommutative geometry,
physical string modes are operators belonging to the
commutant of $\G$ in the crossed product algebra $\AA\sdtimes \G$.
With regard to topology, there is no difference in
the K-theory of the crossed product algebra and that of
its $G$-invariant subalgebra \bratteli.

Let us close this general overview with a remark. In  
 the early days of conformal
field theory (\cf\ \ginsparg) it was noted that for 
some abelian orbifolds one could ``orbifold twice'' and 
recover the original theory. We can explain this 
phenomenon easily in the present context as a result of 
``Takai duality''  (\blackadar, Theorem 10.1.2). 
If  $\G$ is a locally compact abelian group, then
$\AA\sdtimes\G$ has a $\hat\G$ action, where $\hat\G$ is the dual group.
We can therefore gauge the $\hat\G$ symmetry, and in this sense we 
can  ``orbifold twice.'' The Takai duality theorem says that
\eqn\doubleup{
  (\AA \sdtimes \G) \sdtimes \hat\G =  \AA \otimes \KK(L^2(\G))\ ,
}
where $\KK$ is the algebra of compact operators. Thus, 
  ``orbifolding twice'' gives back the original theory,
up to Morita equivalence. The essential physical phenomenon 
is that the twisted sectors from $\hat \G$ restore the states 
which were projected out by $G$.

\subsec{Type IIA}

Let us consider first the type IIA $D9$-brane.
Suppose we have a covariant representation as above;
since $\HH$ is a representation of a locally compact group $\G$
(actually, we are taking it to be finite),
we can decompose $\HH$ into its {\it isotypical components}:
\eqn\decomp{
\HH= \oplus_{\rho\in \widehat{\G}} \HH_\rho\ ,
}
where $\widehat{\G}$ is the set of unitary irreps of $\G$.
Isotypical means the Hilbert space is a direct sum of
copies of the irrep $\rho$.

Again one expects the tachyon field to be 
a projection operator in the large $\Theta$ limit, 
and the solutions to take the form
\eqn\fractbranes{
T = \One - \sum_{\rho\in \widehat{\G}} \P_\rho\ ,
}
where $\P_{\rho}$ is a finite rank projection operator acting
on the fractional brane subspace $\HH_{\rho}$. 
If we think of $\HH_\rho = \HH_0 \otimes \rho$
where $\HH_0$ is some ``standard Hilbert space'' (like
the oscillator space, or more mathematically $\ell^2(\IN)$),
and $\P_\rho$ is of the form $\tilde \P_\rho\otimes 1$ with
$\tilde \P_\rho$ of rank $n_\rho$, then the energy is proportional to
\eqn\energy{
\sum_{\rho\in \widehat{\G}} n_\rho\, \dim(\rho)\ .
}
This solution corresponds to having  a collection of
$n_\rho$ fractional branes of type $\rho$.

More concretely, consider the $d=1$ example of $\IC/\Z_N$.  
According to \automorph, the creation operator transforms as
\eqn\osctransf{
  a\rightarrow \omega a
}
(with $\omega=\exp[2\pi i/N]$) under the generator of $\Z_N$.
Therefore, the number basis states of Hilbert space
transform as 
\eqn\nobasis{
  \ket{n}\rightarrow \omega^n\ket{n}\ ,
}
so that any $n$ of the form $jN+r$ transforms in the $r^{\rm th}$
irreducible representation of $\Z_N$.
The isotypical components are thus
$\HH_r={\rm Span}\bigl\{\ket{jN+r}\bigr\}$ for $j\in\Z$ and $0\le r\le N-1$.
Projectors \fractbranes\ with energy \energy\
have rank $n_r$ in the mod $r$ subspace $\HH_r$ of the
number basis of Hilbert space.

Note that a single unstable $D9$-brane is sufficient to
generate any number of any type of fractional brane in this
simple example.  Equivalently, one could have started with
$N$ unstable $D9$-branes, representing the action of $\G$
as a shift operator in the Chan-Paton space, as in
\dm.  Given that the endpoint of tachyon condensation
is universal, we expect these to give equivalent representations
of $D7$-branes on the orbifold, \cf\ the discussion in section 2.

The ability to represent localized branes on the orbifold 
in terms of tachyon condensation on a single $D9$-brane
is a general feature of orbifolds without discrete torsion,
and follows from the fact that the `oscillator'
ground state $\ket{\vec 0}\in\HH$ may be taken to transform in the
trivial representation of $\G$ (the corresponding 
projector $\ket{0}\bra{0}$ is the image of a gaussian
$2e^{-|z|^2}$ under the map \weyl, and thus invariant
under the point group $\G$).  
The presence of discrete torsion obstructs a construction of
fractional branes using a single unstable $D9$-brane.
{}From \twistcross, the action of $\G$ on the
coordinate operators $a_i$, $a_i^\dagger$ is unchanged,
however the cocycle $\gamma\gamma'=\sigma(\gamma,\gamma')\;\gamma'\gamma$
must be represented on Hilbert space.
The oscillator ground state must therefore lie in 
a projective representation of $\G$ which necessarily
has dimension $M>1$; hence one must start with $kM$ unstable branes.

This requirement of multiple $D9$-branes is an example
of the fact that, if $H$ is $N$-torsion,
then one must have a number of $D9$ branes which is a nonzero
multiple of $N$ \wittenstrings.  In the present case,
$H^2_\G(Y,U(1)) = H^3_\G(Y,\Z)$
for equivariant cohomology groups of a finite group $\G$ acting on Y. 
If we take $Y=\IC^d$ to be contractible, this boils down to
$H^2(\G,U(1)) = H^3(\G,\Z)$; then having a nontrivial
element of discrete torsion means there is an $H$-field turned on
and in our case this is a torsion $H$-field.  

\subsec{Type IIB}

In the $D9$-$\overline{D9}$ system, 
we must construct a set of partial isometries
for fractional branes.  
To begin, let us work with a single brane-antibrane pair. 
Denote by $\Sigma$ a $\G$-invariant shift operator of index $-|\G|$ 
with $\bar \Sigma \Sigma = 1$ (we called this $\bar T$ in equation \teebar,
here we reserve that symbol for the fractional brane partial isometry).  
For example, the $|\G|^{\rm th}$ power of \teebar\ will describe
a regular representation brane at the origin.
Let $\pirho$ be the projection onto the isotypical
component $\HH_\rho$ defined above; then partial isometries
for fractional branes are obtained by projecting the partial
isometry for the regular representation onto fractional brane
subspaces:
\eqn\emilop{
T_\rho = 1- \pirho + \pirho \Sigma \pirho\ .
}
The operator $T_\rho$ acts as a shift operator in
$\HH_\rho$, and as the identity operator in $\HH_{\rho'}$, $\rho'\ne \rho$. 
These operators satisfy a rather interesting algebra:
\eqn\algebra{
\eqalign{
(T_{\rho})^\ell & = 1- \pirho + \pirho \Sigma^{\ell } \pirho \cr
((T_{\rho})^\ell )^\dagger(T_{\rho})^\ell & = 1 \cr
(T_{\rho})^\ell ((T_{\rho})^\ell)^\dagger &  
	= 1 - \pirho \P_\ell \pirho \cr
(T_{\rho})^\ell (T_{\rho'})^{\ell '} & 
	= (T_{\rho'})^{\ell'} (T_{\rho})^\ell 
	= 1- \pirho - \pirhop +
\pirho \Sigma^{\ell } \pirho + \pirhop \Sigma^{\ell'} \pirhop  
\qquad \rho \not=\rho'\ .\cr}
}
Here ${\bf P}_\ell$ is a rank $\ell$ projection operator.
{}From this we can say that the collection of  $\ell_\rho$ fractional
branes of type $\rho$ corresponds to the partial isometry
\eqn\frctpart{
T = \prod_{\rho\in \widehat{\G}} (T_\rho^{\ell_\rho})
}
The energy is proportional to
\eqn\energytwo{
{\Tr}_{\HH}U(\bar T T-1) + {\Tr}_{\HH} U(T\bar T-1) =
U(-1) \sum_{\rho} \ell_\rho \dim(\rho)\ ,
}
where $U(-1)$ is the potential at the unstable maximum.

\bigskip
{\it Example:} 
For the orbifold $\IC/\Z_N$, the shift operator is simply 
$\Sigma=\sum_i\ket{i+N}\bra{i}=
\bigl(a^\dagger\frac1{\sqrt{aa^\dagger}}\bigr)^N$.
The isotypical components of $\HH$ are again the $r$ mod $N$ subspaces
$\HH_r$, as in the previous subsection.  Although the resulting branes
\frctpart\ carry nontrivial charges in K-theory,
and hence are stable in classical open string field theory,
the orbifold breaks supersymmetry.
In the twisted sectors, there are still massless RR fields
coupling to the conserved D-brane charges, but there are also
NS-NS tachyons, so that the closed string vacuum is unstable.
It would be interesting if one could discern the fate
of these charges under closed string tachyon condensation.

\bigskip
{\it Example:} 
The next interesting examples are the ADE orbifolds $\IC^2/\G$
\refs{\dm,\jomy}.
The matrix $\Gamma\cdot x$ of the ABS construction is simply 
$\Bigl({a_1^\dagger~-a_2\atop a_2^\dagger~~a_1}\Bigr)$,
on which the rotation group acts by $SU(2)_L\times SU(2)_R$
transformations.
Supersymmetry and the symplectic form
$\Theta$ are preserved for $\G\in SU(2)_L$, 
and the closed string vacuum is stable.
\item{1.)} {$A_{N-1}$}:
The cyclic group orbifold $\IC^2/\Z_N$ is generated by the element $g$ 
acting on the raising operators as 
\eqn\znorb{
  g(a_1^\dagger,a_2^\dagger)=
	(\omega a_1^\dagger,\omega^{-1}a_2^\dagger)\ .
}
The isotypical components $\HH_r$ of $\HH$ are thus $\ket{n_1,n_2}$
with $n_1-n_2=r$ mod $N$.
\item{2.)} {$D_{N+2}$}:
The dihedral group orbifold $\IC^2/D_{N+2}$ is generated by \znorb\ 
with $\omega=\exp[i\pi/N]$, as well as
\eqn\dnorb{
  h(a_1^\dagger,a_2^\dagger)=(ia_2^\dagger,ia_1^\dagger)\ ,
}
which acts on $\HH$ as $\ket{n_1,n_2}\rightarrow i^{n_1+n_2}\ket{n_2,n_1}$.
The order of the group is $|\G|=4N$.
The isotypical components comprise four sectors  of 
one-dimensional representations 
\eqn\onedreps{\eqalign{
  \HH_0^\pm&={\rm Span}\Bigl\{ \ket{\ell,2jN+\ell} 
		\pm\ket{2jN+\ell,\ell}\Bigr\}
	\quad,\qquad j,\ell\in\Z_+\cr
  \HH_{N}^\pm&={\rm Span}\Bigl\{\ket{\ell,(2j+1)N+\ell}
		\pm\ket{(2j+1)N+\ell,\ell}\Bigr\}
	\quad,\qquad j,\ell\in\Z_+
}}
together with $2(N-1)$ sectors based on two-dimensional
representations
\eqn\twodreps{\eqalign{
  \HH_{r} = {\rm Span}\Bigl\{ 
	\bigl(\ket{\ell,r+2jN+\ell}&,\ket{r+2jN+\ell,\ell}\bigr)\Bigr\}
		\quad,\cr
	&j,\ell\in\Z_+\ ,\quad r=1,...,N-1,N+1,...,2N-1\ .
}}
\item{3.)} {$\TT,\OO,\II$}:
For the group action in the $E_{6,7,8}$ series
(tetrahedral, octahedral, icosahedral subgroups of $SU(2)$), 
the reader may consult for example \jomy.  The decomposition of $\HH$
into its isotypical components is a straighforward if tedious exercise:
\eqn\genreps{
  \HH_\rho={\rm Span}\Bigl\{
	\frac1{|G|}\sum_{g\in G}\chi_\rho(g)\pig(g^{-1})\ket{n_1,n_2}
	\Bigr\}\ .
}

\bigskip
\noindent
The structure of the effective field theory is succinctly
encoded in a quiver diagram \refs{\dm,\jomy}.  For the ADE series,
these are the corresponding extended Dynkin diagrams
(see Figure 1).

\bigskip
{\vbox{{\epsfxsize=4in
    \centerline{\epsfbox{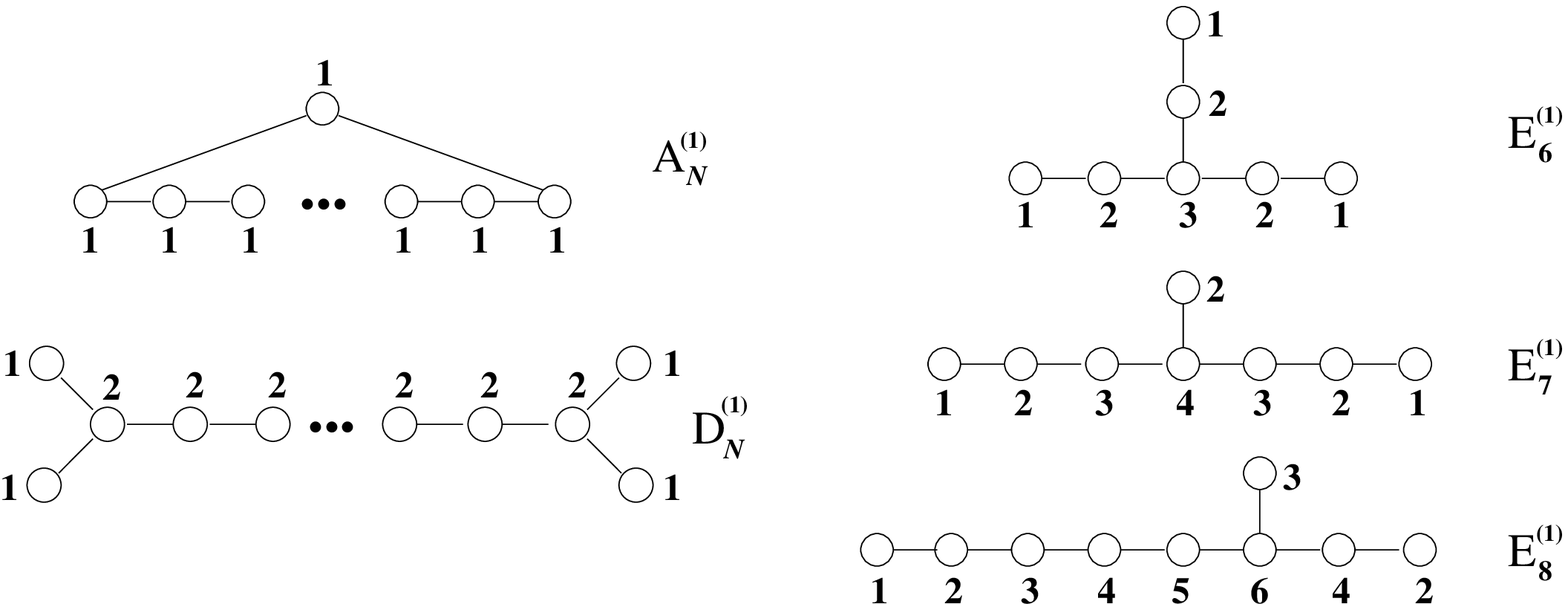}}
    {\raggedright\medskip \vbox{
{\bf Figure 1.}
{\it Dynkin diagrams for affine Lie algebras.  The integers
attached to each node are the Dynkin labels of the affine root.}
 }}}}
    \bigskip}

\noindent
Each irrep is associated to a node of the Dynkin diagram,
and has dimension $d_\rho$ equal to the Dynkin label of that node; 
the $\rho$ representation appears $d_\rho$ times in the regular
representation, and thus leads to $U(d_\rho)$ enhanced gauge symmetry
for the fractional brane at the origin.  
In the present context, the nodes label isotypical components of the
Hilbert space, \eg\ \onedreps\ on the corners of the $D_N$ diagram
and \twodreps\ for the chain of two-dimensional representations.
A general quiver representation consists of $n_\rho$
fractional branes of type $\rho$.
The links of the diagram specify the bifundamental matter content,
which arises from components of the gauge field components
$A_i$ along the orbifold.  A link joining representations $\rho$
and $\rho'$ has multiplicity $n_\rho n_{\rho'}$ in the low energy
Lagrangian.  It is important to note that, 
if $n_\rho$ is zero on some node, then that means the matter
fields $A_z$ in question make open strings
with one end on a fractional brane, and another end `on the closed string
vacuum'.
These are precisely the states that obtained
string scale mass due to the Higgs mechanism in \hklm, 
and hence are not reliably treated in the low energy approximation;
as discussed in section 2.2,
they are argued generally to be removed by stringy effects
not visible at the level of the effective theory.


\subsec{Translating branes away from the orbifold point}

The generic fractional brane has no moduli space corresponding
to translations away from the orbifold point; for instance,
in the BPS case it carries the charges of 
a bound state of branes wrapping various collapsed
cycles at the fixed point \refs{\dm,\ddg,\bcd},
thus it is pinned there.
Indeed, as we just argued, a single fractional brane 
($\ell_\rho=1$, and $\ell_{\rho'}=0$ for $\rho'\ne\rho$)
has no light scalar field in its effective action
that could serve as a Goldstone mode.
However, a brane in the regular representation is free to move away.
Indeed, a brane localized at a generic point
on the orbifold consists of a $\G$-orbit
of such branes on the covering space; the orbit {\it is}
the regular representation on the coordinates.  
Again, the fact that the noncommutative framework reproduces
the correct low-energy dynamics guarantees that the moduli
space will be the same as the commutative description;
one can simply give an expectation value to the transverse
scalars $A_i$ to move a regular representation brane away
from the origin, as in section 2.3.  
Thus, regular representation branes away from the orbifold point
are special cases of \TZ, \generalK\ for points arrayed
along a $\G$-orbit.

To illustrate the presence of a moduli space 
for the regular representation brane,
and the absence of translational moduli for fractional branes,
consider the $\IC/\Z_N$ orbifold.   Implement the orbifold projection
on the algebra $\AA\otimes\matn$ via the projection 
$\OO=\UU\OO\UU^{-1}$, with
\eqn\HN{
  \UU=\pmatrix{ 0&0&\cdots&\cdots &\gamma\cr
                \gamma&0& & &\vdots \cr
                0&\gamma&\ddots& &\vdots\cr
                \vdots& &\ddots&0&0\cr
                0&\cdots &\cdots&\gamma&0}\ ,
}
where $\gamma$ acts in $\HH$ as usual by $\gamma\ket{z}=\ket{\omega z}$.
An invariant tachyon projector for the regular representation 
translated away from the origin is 
\eqn\HNtach{
  \One-T=\pmatrix{ \ket{z}\bra{z}&0&\cdots&0\cr
                0&\ket{\omega z}\bra{\omega z}& &\vdots\cr
                \vdots& &\ddots &0\cr
                0&\cdots&0&\ket{\omega^{-1} z}\bra{\omega^{-1} z}}\ ;
}   
the corresponding gauge field is%
\foot{Note that for $A_z=0$, the projector \HNtach\ is 
unitarily equivalent to the standard projector $T=\One-\P_N$
built from \TZ; 
note also that the physical displacement of the branes 
involves the {\it relative} orientation of the tachyon
and gauge fields in $\matn(\AA)$.}
\eqn\HNA{
  A_z=\pmatrix{z\One&0&\cdots&0\cr
                0&\omega z\One& &\vdots\cr
                \vdots& &\ddots &0\cr
                0&\cdots&0&\omega^{-1}z\One }\ .
}         
Clearly $\Tr|[A_z,T]|^2$ vanishes, and $z$ parametrizes motion
of the brane on the orbifold.  On the other hand,
a generating set of fractional brane projectors is
\eqn\fbproj{
  T_{ab}^{(\ell)}=\coeff1N 
	\omega^{\ell(a-b)}\ket{\omega^a z}\bra{\omega^b z}\ ,
}
for $\ell=0,...,N-1$; here $a,b=0,...,N-1$ are Chan-Paton indices.
For a fractional brane $T=\sum_\ell \vareps_\ell T^{(\ell)}$
with $\vareps_\ell=0,1$, one readily verifies that
\eqn\fracmass{
  \Tr|[A,T]|^2 = |z|^2 \vev{\vec \vareps,\vec \vareps}
}
where $\vev{\ ,\;}$ is the inner product with
respect to the Cartan matrix of the affine Lie algebra ${\hat A}_{N-1}$.    
Thus the attempt to excite transverse motion of fractional
branes costs string scale energy, verfying our claim that
such excitations are among those that should be dropped from
a proper low-energy description.  On the other hand,
superposing all the fractional brane projectors recovers
the regular representation soliton \HNtach\ 
(the off-diagonal elements cancel due to $\Z_N$ phases);
correspondingly, the vector $\vareps_\ell=1$ for all $\ell=0,...,N-1$
is precisely the zero norm vector for the inner product \fracmass,
so that the expectation value of $A_z$ parametrizes a flat direction.


\newsec{Compact orbifolds}

We described in the previous section the ingredients for constructing
D-branes on the orbifold $\IR^{2d}/\G$ in terms of the crossed
product $\AA\sdtimes \G$ of the Moyal algebra of functions $\AA$
on the noncommutative hyperplane $\IR^{2d}$ 
with the orbifold group $\G$.  In fact this prescription provides
a procedure to obtain the D-branes on a general orbifold.  
Given an algebra $\CC(\YY)$ of functions on a 
(possibly noncommutative) space $\YY$,
and carrying a group action $\alpha:\G\rightarrow {\rm Aut}(\CC(\YY))$,
the crossed product $\CC(\YY)\sdtimes\G$ constructs the orbifold.  
The physical string modes are again the commutant of $\G$
in the crossed product.

\subsec{The torus as $\IR^d/\Z^d$}

As a first example, let us construct the algebra of functions 
${\AA}_\theta(\T^2)$ on the two-dimensional noncommutative torus
as the commutant of $\Z^2$ in the crossed
product $\AA(\IR^2)\sdtimes\Z^2$
(this has been discussed in \rieffel, section 2). 
The generators $U$, $V$ of the algebra $\AA_\theta$
satisfy the relation
\eqn\nct{
  VU=e^{2\pi i\theta} UV\ .
}
Let $\TT(m\omega_1+n\omega_2)$
span the algebra of translation operators \transgen\
representing translations along the basic periods of the torus.
Elements of $\AA(\IR^2)\sdtimes\Z^2$ are written
\eqn\genelt{
  \sum_{m,n\in\Z} a_{mn}g_{mn}
}
for $a_{mn}$ in the Moyal algebra of functions on $\IR^2$
and $g_{mn}\in \Z^2$, \ie\ $g_{mn}g_{pq}=g_{m+p,n+q}$.
The crossed product algebra is
\eqn\cpt{
  \sum_{mn}a_{mn}g_{mn}\cdot\sum_{pq}b_{pq}g_{pq}
	=\sum_{mnkl}a_{mn}\alpha_{g_{mn}}(b_{pq})
		g_{m+p,n+q}\ ,
}
where the group $\Z^2$ is taken to act on $\AA(\IR^2)$ 
via conjugation by translations \transgen
\eqn\toract{
   \alpha_{g_{mn}}({\cal O})=\TT(m\omega_1+n\omega_2)
	\,\OO\,\TT(-m\omega_1-n\omega_2) \ .
}
Now let $\omega'_i$ $i=1,2$ be such that 
$\omega'_i\bar\omega_j-\bar\omega'_i\omega_j=2\pi i\delta_{ij}$.
Then from \transgen\
the commutant of $\G=\IZ^2$ in $\AA(\IR^2)\sdtimes\Z^2$ 
is spanned by elements of the form $\TT(p\omega'_1+q\omega'_2)g_{rs}$.
In other words, we identify $U=\TT(\omega'_1)$, $V=\TT(\omega'_2)$,
and $\theta=\omega'_1\bar\omega'_2-\bar\omega'_1\omega'_2$,
and the commutant is $\AA_\theta\times \G$. 
As in all examples of abelian orbifolds,
the commutant of $\G$ in the crossed product is the 
collection of invariant open string states we want,
times a copy of functions on the group (which is
equivalent to Chan-Paton structure).

Since the subalgebra $U^j$ is
commutative, we can alternatively orbifold first by the
action of one $\Z$ factor, and regard the algebra $\AA_\theta$
as $\CC(\T)\sdtimes\Z$.  This means there will be representations
of $\AA_\theta$ on the Hilbert space $L^2(\T)$,
by clock and shift operators.
All these different algebraic descriptions of the noncommutative
torus -- $\AA(\IR^2)\sdtimes\Z^2$, $\CC(\T)\sdtimes\Z$,
and $\AA_\theta$ -- are equivalent \bratteli.

The story for higher-dimensional noncommutative tori is
quite similar \rieffel.
The $d$-dimensional noncommutative torus algebra $\AA_\Theta$ is
generated by translation operators $U_i$, $i=1,\dots, d$,
with relations $U_i^* = U_i^{-1}$ and 
$U_i U_j = \exp[2 \pi i\, \Theta^{ij}]\, U_j U_i$,
where $\Theta^{t} = -\Theta$.  
This algebra will arise as the commutant of the algebra
of translations by a lattice on the noncommutative plane with  
basis vectors $\omega_i$.
The generator of translations is
$\partial_\alpha=-i\Theta^{-1}_{\alpha\beta}[x^\beta,\,\cdot\,]$,
so lattice translations are implemented by
\eqn\latttrans{
  \TT_{\vec n}=\exp[n^i\omega_i^\alpha\partial_\alpha]\ .
}
Repeating the same steps as above, 
the commutant of the $\TT_{\vec n}$ is spanned by
\eqn\Um{
  U_{\vec m}=\exp[2\pi im_j\tilde\omega^j_\alpha x^\alpha]\ ,
}
where $\tilde\omega^j$ are basis vectors for the dual lattice
(momentum space).
Thus the $U_i$ are generators of translations on the lattice
of momenta on the torus.
The algebra $\AA_\Theta$ has a grading by $\Z^d$,
and the vector space basis for the
algebra $U_{\vec m}$, ${\vec m}\in \Z^d$, has multiplication
\eqn\gentoralg{\eqalign{
U_{\vec m} U_{\vec n} &= 
	U_{\vec m+\vec n}\; 
	e^{2\pi i\,m_i\Gamma^{ij}n_j} \cr
  U_{\vec n}U_{\vec m}&= 
	U_{\vec m}U_{\vec n}\; 
	e^{2\pi i\,n_i\Theta^{ij}m_j} \ .
}}
By a redefinition 
$U_{\vec m}\rightarrow\exp[2\pi i\vec m\cdot M\cdot\vec m]U_{\vec m}$,
where $M$ is a symmetric matrix, we can shift away 
the symmetric part of $\Gamma$.
The antisymmetric part is constrained to be
$\Gamma - \Gamma^{t} = \Theta$; 
in particular, one could take
$\Gamma = \half \Theta$.
The general element of the algebra is a linear combination
of the $U_{\vec n}$ with coefficients rapidly decreasing
for large $|\vec n|$. (The $C^*$ algebra is then obtained 
by taking the norm completion.) 

\subsec{String field action} 

Now let us consider the truncation of the string field theory action
which was the starting point for the discussion of \hklm\ in the 
noncompact case. The tachyon fields and hence the 
Lagrangian density should be functions from  $\XX$ to 
the algebra $\AA_{\Theta}$. Moreover, we expect that the Lagrange 
density should have the same form as in \openact. 
Since the grading on the algebra
by $\Z^d$ is a grading by momenta, the action must be given by 
the {\it trace} on the algebra $\AA_\Theta$, defined by 
\eqn\tracealg{
\tau\Bigl(\sum_{\vec n \in \Z^d} a_{\vec n} U_{\vec n}\Bigr) := a_{\vec 0}
}
Therefore, with the simple replacement of the Hilbert space 
trace in \openact\ by the trace on $\AA_{\Theta}$ we obtain 
the action for compactification on the torus (for IIA; IIB is similar):
\eqn\openactcomp{\eqalign{
S = {c\over G_s} \int_\XX  d^{10-d}x \sqrt{G} \,\tau \Bigl[\hf
 & f(T^2-1)G^{\mu\nu} D_\mu T D_\nu T  - V(T^2-1) \cr
 & - {\coeff14} h(T^2-1) F_{\mu\nu}F^{\mu\nu} 
	+\ldots\Bigr]\ .
}}
Here we use a
derivative that acts on the algebra as
$\partial_i U_j=2\pi i\delta_{ij}U_j$,
motivated by the above construction of the torus algebra
from the quotient of the noncommutative plane.
Technically speaking, we should also specify a projective module
(the analogue of a vector bundle for noncommutative algebras)
for which the gauge field provides a connection;
for simplicity, we take the free module $\AA_\Theta^N$
provided by the algebra itself,
corresponding to a topologically trivial connection on the
worldvolume theory of $N$ noncommutative D9-branes.

The action \openactcomp\ arises in the zero-slope
limit (\cf\ \refs{\cds,\sw}) of string dynamics on the torus
in the presence of a $B$ field.  In this limit,
the closed string metric $g_{ij}\to 0$, so closed
string modes of nonzero momentum decouple.
One might worry that the limit is afflicted by a tower
of closed string winding modes descending to 
zero mass; however, the term $\alpha'^2w^i(Bg^{-1}B)_{ij}w^j$
in their mass squared remains finite (and of course, they
only couple to the dynamics at the quantum level).
A somewhat more general noncommutative limit, 
$\alpha'B\to\infty$ at fixed $g$, was considered in \hklm.
One might wonder whether this limit is valid on the torus,
where the moduli space is identified under shifts
of $\alpha'B$ by integers.  The point is that the action
\openactcomp\ refers to a particular (free) module,
and one cannot shift away a large $B$ 
without changing the module to another one with large
lower brane charge (whose physics would of course be
equivalent).  If we fix the brane we have before
tachyon condensation, then the meaning of large $B$
on that brane is unambiguous.



\subsec{D-brane projectors on $T^d$}

According to the action of the previous subsection the
non-BPS branes on the noncommutative torus should
again be described, in the zero-slope or Seiberg-Witten limit,
by projection operators \refs{\schnabl,\bars,\morocco}.
In general, the range of a projection operator $\P$
describing a collection of noncommutative D-brane solitons
is the Chan-Paton space of those branes; 
the algebra $\P\AA\P$ is the algebra of endomorphisms
of the Chan-Paton space.  In the noncompact case
of the noncommutative plane,
this space is finite (say $N$) dimensional, and the 
endomorphism algebra is the algebra of $N\times N$ matrices.
In the torus case, due to the infinite collection of image branes,
the endomorphism algebra should be more complicated.  
Consider the two-dimensional case.  
Start with an unstable D9-brane on a noncommutative
torus associated to the algebra $\AA_\theta$, and condense the tachyon to
$T=\One-\P$; for some appropriate projector $\P$, 
we expect to be able to describe an unstable D7-brane
on the torus.  But by T-duality, the algebra on the D7-brane
must be Morita equivalent to the algebra $\AA_\theta$ we started with --
the T-duality that inverts the torus interchanges D7-branes
and D9-branes, and sends $\theta\rightarrow -1/\theta$
\refs{\cds,\schmo,\mz,\kosc,\pioline,\hv,\sw,\ryang}.
Thus we should have an isomorphism of $C^*$-algebras
$$\P\AA_\theta\P=\AA_{\theta'}, $$
for $\theta'=-1/\theta$.  Let us see that this is so.
First, it is true \raeburn\ that for any $C^*$ algebra $\AA$
and projector $\P$ on $\AA$, $\P\AA$ is an
$\AA\P\AA-\P\AA\P$ {\it Morita equivalence bimodule}.%
\foot{An $\AA-\BB$ Morita equivalence bimodule $\MM$ is
a left $\BB$-module and a right $\AA$-module,
equipped with $\AA$-valued and $\BB$-valued
inner products $\vev{\,\cdot\,,\,\cdot\,}^{ }_\AA$
and $\vev{\,\cdot\,,\,\cdot\,}^{ }_\BB$ such that
$\langle f,g\rangle^{ }_\BB h=f\langle g,h\rangle^{ }_\AA$,
for $f,g,h\in\MM$.
The existence of the bimodule establishes the
Morita equivalence of $\AA$ and $\BB$; see for instance \raeburn.}
Now note that $\AA\P\AA$ is a two-sided ideal of $\AA$.
However, $\AA_\theta$ is a simple algebra
\simplealg, meaning it has
no proper two-sided ideals; therefore 
$\AA_\theta\P\AA_\theta=\AA_\theta$.
We conclude that $\P\AA_\theta\P$ and $\AA_\theta$
are Morita equivalent.  It is a fact \rican\ that
the only $C^*$ algebras with identity that are Morita
equivalent to $\AA_\theta$ are the algebras
$M_n(\AA_{\theta'})$, where $\theta'$ is
related to $\theta$ by a fractional linear transformation,
$\theta'=\frac{a\theta+b}{c\theta+d}$.
Thus the endomorphisms of the Chan-Paton bundle on
the lower brane constitute a noncommutative torus
algebra T-dual to the original one.
To complete the identification, we need only to determine 
the solitonic D-brane numbers of a given projector
via the above T-duality relation.

Projection operators in the two-dimensional 
noncommutative torus algebra
can be given explicitly \refs{\ripo,\boca}.  
Assume (without loss of generality) $\half<\theta<1$, and let
$f$ and $g$ be periodic functions such that
\eqn\torprojfns{\eqalign{
  f(e^{2\pi ix})&=\cases{
		{\rm a~smooth~function~increasing~from~0~to~1}\qquad 
			&$x\in[0,1-\theta]$\cr
		1\hfill &$x\in[1-\theta,\theta]$\cr
		1-f(e^{2\pi i(x-\theta)}) &$x\in[\theta,1]$}\cr
	&\cr
  g(e^{2\pi ix})&=\cases{
	0 &$x\in[0,\theta]$\cr
	\sqrt{f-f^2} &$x\in[\theta,1]$}
}}
Then 
\eqn\torproj{
  \P_\theta=g(V)U+f(V)+U^{-1}\bar g(V)
}
is a projector (constructed by Powers and Rieffel \ripo) which, 
together with the trivial projector $\One$,
generates $K_0(\AA_\theta)$ when $\theta$ is irrational.
The use of this projector to construct D-branes on tori
was discussed in \refs{\schnabl,\bars,\morocco}.  

Projection operators enable us to construct lower-dimensional
non-BPS branes as solitons in higher-dimensional non-BPS branes.
Solitonic field configurations on the torus carrying RR charge can also
be constructed from projection operators.
The discussion involves yet more machinery of 
noncommutative geometry, and is therefore deferred to \paperii.
We will see there that the Powers-Rieffel projector \torproj\ 
may be used to construct the full K-theory lattice of brane bound
states on the two-dimensional torus.
In anticipation of this development (and to explore more fully
the possible unstable brane configurations), 
we recall a construction of a complete set
of projectors $\P_{n+m\theta}$ by Rieffel \ripo.  
Note that a projector 
for $Dp$-$D(p-2)$ charges $(r,s)$ with $0<r+s\theta < 1$ 
is built as follows \ripo:
Let $\CC(\T)$ be the functions on the circle.
The Powers-Rieffel projector \torproj\ uses such functions to construct
the projector for $(r,s)=(0,1)$ or $(1,-1)$ depending on
conventions, with $\theta<1$.  But $\CC_m(\T)$,
the functions of period $1/m$, is a subalgebra of $\CC(\T)$,
and on $\CC_m(\T)$, a shift by $\theta$ (the action of $V$
in the noncommutative torus algebra \nct)
looks like a shift in $\CC(\T)$ by $\{m\theta\}$,
the fractional part of $m\theta$.  Now just repeat the
construction of the projector \torproj\ using this function space;
if the original projector associated to $\theta$
had, say, quantum numbers $(r,s)=(0,1)$, 
then the new one has quantum numbers $r=n$, $s=m$, 
such that $0<n+m\theta<1$.
One can add Chan-Paton structure, put this projector in the
first diagonal entry, and the trivial projector $\One$
in the rest, to get a projector with any $n+m\theta>0$.
These are precisely the stable (real, as oppose to virtual) bundles
in the K-group of the noncommutative 2-torus, which is
isomorphic to $\Z+\Z\theta$ for irrational $\theta$.
The above construction is unique up to isomorphism; it is a theorem
\rieffel\ that (in any dimension), for $\Theta$ irrational, 
any two projective modules
representing the same element of K-theory are isomorphic.

We can now canonically associate particular projectors
$\P_{n+m\theta}$ to the appropriate collection
of unstable branes with brane numbers $(m,n)$.%
\foot{Note that in the IIA case, the K-theory classes
of the projectors $\P_{n+m\theta}$ are not measuring
conserved charges of the brane configuration.
This is because there is a continuous path of
nonsingular configurations 
$T=s\P_{n+m\theta}+(1-s)\P_{n'+m'\theta}$, $0\le s\le 1$,
that smoothly interpolates between any two of them.
Instead, in this case $n,m$ characterize
the various critical points of the action.}
Recall that the normalization of the trace transforms as
\refs{\cds,\schmo,\mz,\kosc,\pioline,\hv,\sw,\ryang}
\eqn\trtrans{
  \tau^{ }_{\theta'}[\quad]=
	(c\theta+d)^{-1}\;\tau^{ }_{\theta}[\quad]
}
under the Morita equivalence that maps
$\theta\rightarrow \theta'=\frac{a\theta+b}{c\theta+d}$.
Since $\tau(\P_\theta)=\theta$, we conclude that 
the trace of the identity in $\AA_{\theta'}$ 
is correctly reproduced by the trace of $\P_\theta$ if 
the algebra $\AA_{\theta'}=\P_\theta\AA_\theta\P_\theta$
has $\theta'=-1/\theta$; indeed this is the T-duality
that maps D7-branes to D9-branes.
Similarly, for the algebra on the range 
$\AA_{\theta''}=(\One-\P_\theta)\AA_\theta(\One-\P_\theta)$,
one finds $\theta''=\frac{-1}{1-\theta}$, which correctly
maps $(D7,D9)$ charges $(-1,1)$ to $(0,1)$.
More generally, $\AA_{\hat\theta}=\P_{n+m\theta}\AA\P_{n+m\theta}$
is associated to brane numbers $(m,n)$ via 
the fractional linear transformation 
$\hat\theta=\frac{a\theta+b}{m\theta+n}$ that transforms
the brane numbers to $(0,1)$.

The potential energy term in \openactcomp\ is proportional to 
the trace \tracealg\ of the projector $\tau(\P_{n+m\theta})$,
and this is the leading term in the energy in the zero-slope or 
Seiberg-Witten limit.%
\foot{The trace gives a measure of the dimension of the 
space $\P_{n+m\theta}\AA_\theta\P_{n+m\theta}$ thought of as an
$\AA_\theta$ module, as well as the brane tension; \cf\ \refs{\cds,\hkms}.} 
A simple calculation yields $\tau(\P_{n+m\theta})=n+m\theta$.%
\foot{The trace is a cohomological invariant on the algebra,
and it provides the map from the 
lattice of K-theory charges to the ordered group 
$\IZ+\IZ\theta$ mentioned above.}
Thus $N$ unstable D9-branes can decay into any ensemble
of unstable branes such that the corresponding projector has
trace $n+m\theta<N$.
In principle, this might lead to an infinite collection
of arbitrarily finely spaced critical points in the tachyon
potential, for instance of a single unstable D9-brane, 
corresponding to successively better
rational approximations $-m/n\sim\theta$.
Considerations of the kinetic energy terms in \openactcomp\
could reduce the number of critical points to a finite number.
Although these critical points are closely spaced in energy,
nearby ones may correspond to widely different field 
configurations (rather similar to the irrational axion \irrax),
with a corresponding large potential barrier.
Each of these configurations is of course unstable toward
decay to the closed string vacuum.

For orbifold applications, $\P_\theta$ can only be used for
$\Z_2$ quotients, since its invariance group is
$U\rightarrow U^{-1}$, $V\rightarrow V^{-1}$;
the Powers-Rieffel projector \torproj\  is ill-suited
for the construction of more general orbifolds which act
by permuting the generators $U$ and $V$ (see below).
A general procedure to construct projection operators
on $\AA_\theta$ uses Morita equivalence. 
We begin with the projective modules for $\AA_{\theta}$ 
introduced by Connes in \connescrasp.  In the simplest 
example the module is the Schwarz space $\SS(\IR)$ 
of smooth functions of rapid decrease
at infinity.  In fact, this module
provides a Morita equivalence bimodule $\MM$ between
$\AA=\AA_\theta$ and $\BB=\AA_{1/\theta}$ as follows
\refs{\connescrasp,\rican}.
Let $VU=\lambda UV$, $\lambda=e^{2\pi i\theta}$, and
$\widetilde V\widetilde U=\mu \widetilde U \widetilde V$, 
$\mu=e^{2\pi i/\theta}$.  
We have left and right actions of $\BB$ and $\AA$, respectively, 
on $f\in\SS(\IR)$ via
\eqn\aact{\eqalign{
  (fV)(t)=e^{2\pi it}f(t)\qquad~ &,\qquad
  (fU)(t)=f(t+\theta)\cr
  (\widetilde Vf)(t)=e^{-2\pi it/\theta}f(t)\quad &,\qquad
  (\widetilde Uf)(t)=f(t+1)\ .
}}
Then for functions
$f,g\in\SS(\IR)$ we can define $\AA$- and $\BB$-valued inner products
\eqn\aprod{\eqalign{
  \langle f,g\rangle^{ }_\AA&=\sum_{m,n} \langle f,g\rangle^{ }_\AA(m,n)
	\cdot U^mV^n\cr
  \langle f,g\rangle^{ }_\AA(m,n)&=\theta\int_{-\infty}^\infty
	\overline{f(t+m\theta)}g(t)e^{2\pi i(-nt)}\;dt
}}
and
\eqn\bprod{\eqalign{
  \langle f,g\rangle^{ }_\BB&=\sum_{m,n} \langle f,g\rangle^{ }_\BB(m,n)
	\cdot \widetilde U^m\widetilde V^n\cr
  \langle f,g\rangle^{ }_\BB(m,n)&=\int_{-\infty}^\infty
	f(t-m)\overline{g(t)}e^{2\pi i(nt/\theta)}\;dt\ .
}}
One can show (\rieffel, section 2) that 
\eqn\twoprods{
\langle f,g\rangle^{ }_\BB h=f\langle g,h\rangle^{ }_\AA,
}
which is the key statement of Morita equivalence.  
In particular, the norm completion of 
$\MM_{ab}\nobreak\otimes_\BB\nobreak\MM_{ba}$ is $\AA$,%
\foot{The notation $\MM\otimes_\BB\NN$ stands for the 
bimodule spanned by elements of the form $m\otimes n$ subject 
to the relations that
$mb\otimes n=m\otimes bn$ for all $m\in\MM$, $n\in\NN$, $b\in\BB$.}
and of $\MM_{ba}\otimes_\AA\MM_{ab}$ is $\BB$.%
\foot{In string terms, this means that sewing open
strings with boundary conditions $ab$ and $ba$ 
yields the full algebra of $aa$ strings or $bb$ strings, depending
on which ends are sewn. The fact that  $\MM_{ba}\otimes_\AA\MM_{ab}$
is the entire algebra $\AA_{bb}$ follows from the 
generalized Cardy condition.}
One constructs nontrivial modules by picking suitable 
functions $f_i$, $i=1...N$
such that $\sum_i\langle f_i,f_i\rangle^{ }_\BB=\One_\BB$;
then $\langle f_i,f_j\rangle^{ }_\AA$ is a nontrivial projector
in $\matn(\AA)$.%
\foot{The module just constructed is that of $p$-$(p-2)$ strings, \cf\ 
\refs{\mz,\sw}.
Thus it naturally relates the trivial projector for, say, the
algebra of $(p-2)$-$(p-2)$ strings, to a nontrivial projector for
the algebra of $p$-$p$ strings.  Indeed, Poisson resummation
is a key ingredient both in the demonstration \rieffel\ of 
Morita equivalence of $\AA_\theta$ algebras and in string T-duality.}
In particular, \boca\ constructs a projection
operator homotopic to \torproj\
starting from the Schwarz function $f=e^{-\pi t^2/\theta}$. 
It turns out (and this is nontrivial) that  
$B=\langle f,f\rangle^{ }_\BB$ is invertible. It 
then easily follows that   
\eqn\bocaproj{
  \P_\theta=\langle B^{-1/2}f, B^{-1/2}f\rangle^{ }_\AA
}
is a projector.  Since the projector is built out of
the gaussian function, it is clearly invariant under
the $\Z_4$ operation of Fourier transformation in $\SS(\IR)$,
which sends $U\rightarrow V$, $V\rightarrow U^{-1}$.
For special rational values $\theta=1/q$, $q\in\IZ$,
one can find explicit expressions for $\P_\theta$
in terms of theta functions of $U$ and $V$ \boca.
We give a general expression below. 

The above elegant constructions, due to Connes and Rieffel 
\refs{\connescrasp,\rican,\rieffel},
of the projective modules and their Morita equivalence 
properties generalizes beautifully to higher dimensional tori,
in terms of a representation on 
$L^2(\IR^p\times\Z^q\times F)$, 
where $F$ is a finite group and $2p+q=d$.  
We give a brief summary of it in appendix A.
(This construction is also reviewed in \koscrev,
and interpreted physically in \sw.)  


Since there are exact solutions
\solgen\ for the noncommutative plane, 
one might hope that one can 
find a compatible gauge field such that the 
tachyon and gauge field equations of motion 
on the torus are
solved beyond leading order in the limit
of large noncommutativity. We will now give 
a partial solution to this problem. 
The tachyon field equations will be solved exactly
if we can find a compatible connection such that $DT=0$.
The above bimodule construction of projectors
is helpful in this regard by allowing us
to find such a connection.
Some useful identities for the bimodule $\MM$ are
\eqn\modprops{\eqalign{
  \langle f, g\rangle_{\AA}^* = 
	\langle g, f\rangle^{ }_{\AA}\qquad&\qquad 
  \langle f, g\rangle_{\BB}^* = 
	\langle g, f\rangle^{ }_{\BB} \cr
  \langle f, g a\rangle^{ }_{\AA} = 
	\langle f, g\rangle^{ }_{\AA} \;a\qquad&\qquad
  \langle b f, g\rangle^{ }_{\BB} =
	b\; \langle f, g\rangle^{ }_{\BB} \cr
  \langle fa^*, g\rangle^{ }_{\AA} = 
	a\; \langle f, g\rangle^{ }_{\AA} \qquad&\qquad
  \langle f, b g \rangle^{ }_{\BB} = 
	\langle f, g\rangle^{ }_{\BB} \;b^*\ ,
}}
where $f,g\in\MM$, $a\in\AA$, $b\in\BB$.
The derivative acts as (for $\Phi=-\Theta^{-1}$)
\connescrasp%
\foot{Warning: There are some incompatible 
factors of $2\pi$ between 
standard conventions for the noncommutative plane and torus,
\cf\ \ncder.}
\eqn\torder{
\eqalign{
(d_1 f)(t) & = - {2\pi i \over \theta} t f(t) \cr
(d_2 f)(t) & = {d \over dt}  f(t) \ ;
}}
this reproduces the derivation $\p_i U_j=2\pi i\delta_{ij}U_j$
on the algebra provided we identify 
\eqn\bimodder{
\eqalign{
  \p_i (\langle f, g\rangle_{\AA})  & = \langle d_i f, g\rangle_{\AA} + 
	\langle f, d_i g\rangle_{\AA} \cr
  \p_i (\langle f, g\rangle_{\BB})  & 
	= - \theta \bigl(  \langle d_i f, g\rangle_{\BB} + 
	\langle f, d_i g\rangle_{\BB} \bigr) \ .
}}
Note that from the bimodule property
$\langle  f, g\rangle_{\BB} \;h  =f\; \langle g, h\rangle_{\AA} $,
if there exists $\ff$ with \hbox{$\langle \ff, \ff\rangle_\BB =\One$}
and $\langle \ff,\ff\rangle_\AA = \P$, 
then $h = \ff\langle \ff, h\rangle_{\AA}$ 
for all $h$, in particular for $h=d\ff$. 
Then it is straightforward to verify that
\eqn\Pders{
\eqalign{
\P \langle d \ff, \ff\rangle_{\AA} 
	&=-\langle \ff, d\ff\rangle_{\AA}\P\cr
\langle d \ff, \ff\rangle_{\AA} \P 
	&=\langle d \ff, \ff \rangle_{\AA} \cr
\P \langle  \ff, d \ff\rangle_{\AA} 
	&=\langle\ff,d\ff\rangle_{\AA}\ ,\cr}
}
from which it follows that
$ d\P + A\P - \P A = 0 $
for the connection
\eqn\compatconn{
A =  \langle \ff, d\ff \rangle_{\AA} 
	- \langle d\ff , \ff \rangle_{\AA} \ .
}
Note that, because it is not constructed using
this bimodule procedure, it is not clear whether
the Powers-Rieffel projector \torproj\ 
admits a compatible connection such that the 
tachyon field equations are solved exactly;
therefore, it may only be a solution to
the field equations in the leading order
of the limit of large $\alpha' B$.

It remains to find the function $\ff$ among the class
with $\langle \ff,\ff\rangle_\BB=\One$ that minimizes
the gauge field kinetic energy. 
This is the step we have not carried out. 
The gauge kinetic energy in \openact\ is proportional to
\eqn\ymke{
  \tau\left[\P\,\bigl(dA+[A,A]\bigr)^2\right]\ .
}
We have not succeeded in minimizing this energy. 

As an aside, a rather explicit formula 
for the projector \bocaproj\ may be given.
Let $f(t) = e^{-a t^2 - bt} $ with $a$ real and positive. 
The $\BB$ algebra acts as
\eqn\simpleact{
  (\tilde U^m \tilde V^n f)(t)= 
	f(t) e^{-(2ma + {2\pi i n\over \theta})t} 
	e^{-am^2-bm- {2\pi i m n\over \theta}}
}
Thus, a gaussian goes into a gaussian 
up to a linear exponential in $t$ and a prefactor. 
Thus if we define $B := \langle f,f\rangle_{\BB}$ then 
(simplifying by taking $b=0$)
\eqn\explicit{
B^{-1/2} f = \Biggl( 1+ \sum_{k=1}^\infty 
	{\Gamma(k-1/2)\over \Gamma(-1/2)k!}
	\bigl({\pi \over 2 a}\bigr)^{k/2}
	{\sum_{m_k,n_k}}' e^{-Q_k(m,n)} \prod_{i=1}^k 
	e^{-(2m_i a + 2 \pi i n_i/\theta)t} \Biggr) e^{- a t^2}
}
where the sum $\sum'$ is over tuples 
$(m_1, n_1),\dots,(m_k,n_k)$ 
of nonzero pairs of integers $(m,n)$ and $Q_k(m,n)$ is a quadratic form: 
\eqn\quadrtc{
Q_k(m,n) = {a\over 2} (\sum_{i=1}^k m_i^2) + a(\sum_i m_i)^2
 + {\pi^2 \over 2 a \theta^2}(\sum_{i=1}^k n_i^2) + 
 {i \pi \over \theta}(\sum_i n_i m_i + 2 \sum_{i>j}n_i m_j )\ .
}
%


\subsec{$T^d/G$: torus orbifolds}

The torus is the orbifold of the plane by a lattice,
and the construction of D-brane solitons as
projection operators parallels the noncompact case.
If the Narain data of a torus is left invariant
by some subgroup $\TT_\fix$ of the T-duality group $\TT=O(d,d;\Z)$
of a $d$-dimensional noncommutative torus, one may
the further orbifold by $\TT_\fix$.
The T-duality group $\TT$ consists of transformations 
\eqn\tduality{
  \pmatrix{\aa&\bb\cr \cc&\dd}\quad ,\qquad \aa^t\cc+\cc^t\aa=0~,
	~~ \bb^t\dd+\dd^t\bb=0
	~,~~ \aa^t\dd+\cc^t\bb=\One\ ,
}
acting on the closed string
Narain data $E=g+{2\pi\alpha'}B$ via
\eqn\modtrans{\eqalign{
  E &~~\longrightarrow~~ 
	\bigl(\aa E+\bb\bigr)
	\bigl(\cc E+\dd\bigr)^{-1}\cr
  g_s&~~\longrightarrow 
	~~g_s\sqrt{\det\bigl[\cc E+\dd\bigr]} \ .
}}
In terms of the noncommutative parametrization 
of the Narain data \refs{\kosc,\pioline,\hv,\sw,\ryang}
\eqn\opcldata{
  \frac1{g+B}={\Theta}
	+\frac1{G+\Phi}\ ,
}
the T-duality transformation \modtrans\ 
is associated to the Morita equivalence of the algebra 
$\AA_\Theta$, equation \gentoralg,
to its images under $O(d,d)$ transformations
\eqn\morita{\eqalign{
  \Theta&~~\longrightarrow~~ (\cc+\dd\Theta)(\aa+\bb\Theta)^{-1}\cr
  G&~~\longrightarrow~~ (\aa+\bb\Theta)G(\aa+\bb\Theta)^t\cr
  \Phi&~~\longrightarrow~~(\aa+\bb\Theta)\Phi(\aa+\bb\Theta)^t+
	\bb(\aa+\bb\Theta)^t\cr
  G_s&~~\longrightarrow~~ G_s\;\bigl({\det[\aa+\bb\Theta]}\bigr)^{1/2}\cr
  \tau[\quad]&~~\longrightarrow~~ 
	\bigl({\det[\aa+\bb\Theta]}\bigr)^{-1/2}\;\tau[\quad]\ .
}}                                                          
Automorphisms of the algebra $\AA_\Theta$
require $\bb=0$, so that $\Theta=(\cc+\dd\Theta)\dd^t$; 
they act on the basis elements $U_{\vec n}$ as
$U_{\vec n}\rightarrow\chi(\vec n)U_{\dd^t\vec n}$,
where $\chi(\vec n)\in U(1)$.
It is known that $\Theta$ can be skew-diagonalized over the
integers.  If the skew eigenvalues are not rationally related,
and $d$ is even, then the automorphism group
is precisely $SL(2,\Z)^{d/2}\sdtimes \tilde T^d$.  

The noncommutative torus algebra $\AA_\Theta$ appears
as a decoupled factor in the full string vertex algebra 
$\AA_\str$ only in the zero slope limit 
$\alpha'\rightarrow 0$ \wittach; away from this limit,
one ought to consider $\AA_\str$ itself and carry on the
discussion at the level of string field theory.
In the zero slope limit, 
there is still an action of the T-duality group 
on $\AA_\Theta$ via \morita.
Furthermore, the stringy modes decouple from low-energy physics,
leaving a noncommutative field theory built on $\AA_\Theta$.
In terms of the Narain data, one is approaching the boundary 
of the moduli space where effectively 
$(G+\Phi)\rightarrow\infty$;
the T-duality group acts (ergodically) 
via \morita.  So for orbifolds in which the volume remains as
a modulus (symmetric orbifolds in particular), 
we can discuss the construction of D-branes
as solitons consistently at the level of noncommutative field theory.
Asymmetric orbifolds often fix the Narain data
to some enhanced symmetry locus in the middle of moduli space, 
in which case this freedom is not available.
For the rest of this subsection, we will restrict attention
to orbifolds which allow a decoupling of
the zero mode algebra to $\AA_\Theta$, and discuss the
orbifold action there.  The next subsection deals
with asymmetric orbifolds.

Consider then symmetric orbifolds.  
The orbifold is again associated to a crossed product algebra
$\AA_\Theta\sdtimes\TT_\fix$
\koscorb\
\eqn\torcross{\eqalign{
  U_{\vec m}U_{\vec n}&=e^{2\pi i\vec m\cdot\Theta\cdot\vec n}\,
	U_{\vec n}U_{\vec m}\cr
  \pig(g)U_{\vec n}\,\pig(g)^{-1}&=
	e^{i\chi(\vec n,g)}\,U_{R(g)\vec n} \cr
  \pig(g)\pig(h)&=\sigma(g,h)\,\pig(h)\pig(g)\ ,
}}
where we have included
the possibility of $H$-torsion in the form
of cocycles $\chi$, $\sigma$ in the action of $\G=\TT_\fix$;
$\chi,\sigma$ can also arise when the orbifold group 
includes shifts on the noncommutative torus in addition to rotations,
\cf\ the discussion at the beginning of this section.
Alternatively, one can take the orbifold of the noncommutative
plane $\IR^d$ by the full space group of the orbifold.
For symmetric orbifolds, the orbifold group $G$ consists of
crystallographic symmetries of the lattice defining
the torus.  These are the $SL(d)$ automorphisms defined above, 
together with shifts on the torus dual to the lattice.  
If we further demand invariance of the Hamiltonian,
we are restricted to rotations $O(d)\subset SL(d)$.

The simplest examples to consider are orbifolds 
by symmetric shifts.  From the discussion in section 4.1, 
the elements $U_{\vec n}$
are to be thought of as translations on the lattice of momenta 
$\dlat\cong\Z^d$ on the torus.  
One can think of the lattice $\dlat$ as embedded in the
noncommutative plane $\IR^d$ parametrizing translations \transgen,
and $\Theta$ as determined by the basis vectors of the lattice
and the cocycle for translations in $\IR^d$.
A commutative torus
would be defined by the dual lattice $\dlat^*$ as $\T^d=\IR^d/\dlat^*$;
then an order $N$ symmetric shift is a vector $v$ such that
$Nv\in\dlat^*$ is primitive.   The orbifold by this symmetric shift
decreases the volume of the torus by a factor $N$,
and correspondingly increases the size of the lattice of momenta
by eliminating all $p\in\dlat$ such that $p\cdot v\not\in\Z$.
It is this last fact that tells us how to implement the 
orbifold of the noncommutative torus by a symmetric shift.
Recall that the physical string modes live in the commutant
of $\G$ in the crossed product algebra; thus we simply need
a group action that leaves invariant 
the same basis elements of $\AA_\theta$
as in the commutative case.  It is sufficient to take
$\G=\Z_N$, with generator $g$ acting on $U_{\vec n}$ as
\eqn\symorbact{
  \pig(g)U_{\vec n}\,\pig(g)^{-1}=
	e^{2\pi i\,\vec n\cdot\vec v}\,U_{\vec n}
}
(a special case of \torcross), where again $Nv\in\dlat^*$.
The commutant of $\G$ in $\AA_\Theta\sdtimes\G$ is
the noncommutative algebra of the orbifolded torus.

Now we consider rotations.
A major difference between the compact and noncompact
orbifolds is the presence of multiple fixed points
in the compact case.  While the Hilbert space $\HH$
breaks up into isotypical components under the action of $G$,
this does not reveal the entire fractional brane structure
of the orbifold (roughly speaking, the isotypical components
should only give the fractional brane structure of one
of the fixed points).  In addition to the obvious projectors
\eqn\isotypproj{
  \frac1{|G|}\sum_{g\in G}\chi_\rho(g^{-1})\pig(g)
}
onto isotypical components, there may be other embeddings
$\vareps:\G\rightarrow\AA_\Theta\sdtimes\G$
of the orbifold group into the crossed product algebra.
Let the image of $g\in\G$ in the crossed product algebra be 
$\vareps(g)\,g$; then
\eqn\anotherproj{
  \frac1{|G|}\sum_{g\in G}\chi_\rho(g^{-1})
	\pia(\vareps(g))\pig(g)
}
is a distinct projector.
The projectors \anotherproj\ fall into 
a finite number of unitary equivalence classes
(equivalent projectors reduce in the commutative case 
to fractional branes at covering space images 
of the same fixed point on the torus).
Such projectors typically do not exhaust 
the set of fractional brane projectors of the torus
orbifold, as there can be fixed loci consisting
of orbits of points stabilized by some subgroup of $\G$ (see below).

Let us now turn to examples.
Specialize again to the two-dimensional situation, and suppose that
$\G$ does not contain shift vectors.
Automorphisms of $\AA_\theta$ are generated by $SL(2,\Z)$ 
transformations mixing $U$ and $V$: $(U,V)$ and $(U^aV^b,U^cV^d)$
generate the same noncommutative torus algebra \nct,
provided \hbox{$ad-bc=1$}.  Rotations involved in 
symmetric orbifolds are in
$O(2)\bigcap SL(2,\Z)$
\eqn\twodorbs{\eqalign{
  &\Z_2~:\qquad U\rightarrow U^{-1},\qquad V\rightarrow V^{-1}\cr
  &\Z_3~:\qquad U\rightarrow V\quad,\qquad V\rightarrow U^{-1}V^{-1}\cr
  &\Z_4~:\qquad U\rightarrow V\quad,\qquad V\rightarrow U^{-1}\cr
  &\Z_6~:\qquad U\rightarrow V\quad,\qquad V\rightarrow U^{-1}V\ ,
}}
While {\it any} subgroup of $SL(2,\Z)$ can serve as an automorphism 
group on $\AA_\theta$, we restrict attention to those groups which 
preserve the Hamiltonian. It turns out that these are the same as the 
finite subgroups of $SL(2,\Z)$.  

The general $SL(2,\Z)$ transformation acts unitarily on $\HH$.
As in the noncompact case, the Hilbert space breaks up into $|\G|$
isotypical components under the action of $\G$, and
one might imagine proceeding as in the previous section.
Alternatively, a regular representation brane may be constructed 
by embedding a given projector, such as \torproj, 
in a $|\G|$-dimensional Chan-Paton space 
with successive entries rotated
according to the $\G$ action \twodorbs\ (\cf\ \HNtach).
For a different construction of modules of the crossed product
in the $\Z_2$ case, see \koscorb.
In the $\Z_4$ orbifold, in addition to
the three nontrivial projectors \isotypproj\ one has
three more of the form \anotherproj\ using 
$\vareps(g)=e^{i\pi\theta/2}\,U$,
where $g$ is the generator of $\Z_4$ \walters.
In the commutative setting ($\theta=0$), $\vareps(g)g$ represents
a $\pi/2$ rotation followed by a lattice translation,
so we might think of this second set of projectors
as related to pointlike branes on the orbifold concentrated
at the `point' on the torus fixed by this transformation
rather than at the `origin' in the covering space fixed by $g$;
of course, this is rather imprecise language, since there
are no `points' in noncommutative geometry.
In the $\Z_4$ example, the six projectors discussed above exhaust 
the set of independent projectors of the form \anotherproj\ \walters.%
To complete the list of such projectors for two-dimensional
toroidal orbifolds, for $\Z_2$ one has four projectors
based on $\vareps(g)=1,U,V,e^{i\pi\theta}UV$;
for $\Z_3$, one has six projectors based on $\vareps(g)=1,U,V^{-1}$
(two each for two independent choices of a cube root of unity);
and for $\Z_6$ there are only the five nontrivial projectors \isotypproj\
onto the isotypical components of $G$.

Projectors of the type \anotherproj\ characterize pointlike
branes on the orbifold.
In the $\IZ_2$ case, one need only add the projector $\P_\theta$
\torproj\ or \bocaproj\ to span the lattice of K-theory charges.
In the $\IZ_4$ case, one must use the $\IZ_4$ invariant
projector \bocaproj;
three further projection operators, needed to span the lattice
of K-theory charges, arise from \bocaproj\ for different
choices of the action of the generator 
$g$ of $\Z_4$ on the module $\SS(\IR)$ --
$gf=\hat f$, $i\hat f$, or $-\hat f$ for $f\in\SS(\IR)$,
where $\hat f$ is the Fourier transform of $f$.  
These describe two different branes associated to orbits of
`points' fixed by $\Z_2\subset\Z_4$, but not by $\Z_4$, as well as
the `untwisted sector' brane described by \bocaproj\
before orbifolding.
For $\Z_3$, one needs only one nontrivial projector of the 
type \bocaproj, but invariant under the $\IZ_3$ action \twodorbs; 
for $\Z_6$, to completely span the lattice of K-theory charges
requires four additional such projectors.
We have not found explicit expressions for the
projectors in these two cases.

\subsec{Asymmetric orbifolds}

A very interesting   set of backgrounds 
are   asymmetric  toroidal orbifolds. These are obtained by 
gauging a subgroup $\TT_\fix$ 
of the T-duality group $O(d,d)$ that leaves the Narain data fixed.
In general this is only possible at 
  enhanced symmetry points, which  are in the middle of the
Narain moduli space. Therefore,  we can no longer simplify the
analysis by taking the zero slope limit -- 
the metric and $B$-field are necessarily of order the string scale.
As a simple example, consider an orbifold by $T$-duality itself. 
In this case $E \to E^{-1}$ preserves the Narain data only 
when $B=0$ and the size of the torus is string scale. 
Evidently, one can neither take the limit of \sw\ nor of \hklm!
A discussion of branes in asymmetric orbifolds along the 
above lines must therefore use the full  
noncommutative structure of the 
string field algebra. 
There is also typically no advantage to working with the
noncommutative description   in terms of $G$, $\Theta$, and $\Phi$.
The noncommutativity inherent in the string scale will
be of the same order as that due to $\Theta$;
furthermore, the noncommutative data transform inhomogenously \morita. 
Transformations which fix a point $E=g+B$ in Narain
moduli space generically will not fix the noncommutative
description, thereby obscuring the presence of an enhanced
symmetry.  Therefore, we proceed 
formally with the standard open string vertex algebra
in its usual description in terms of background data $g$, $B$,
and make a proposal for how to formulate 
D-brane projectors. 

We would like to build an algebra on which $\TT_\fix$
acts as a group of automorphisms.  
A natural object in this regard is
\eqn\Balg{
  \BB=\pmatrix{\AA_{a_1a_1}&\MM_{a_1a_2}&
		\cdots&\MM_{a_1a_n} \cr
	\MM_{a_2a_1}&\AA_{a_2a_2}& &\vdots \cr
	\vdots& &\ddots& \cr
	\MM_{a_na_1} & \ldots& &\AA_{a_na_n}}\ ,
}
the {\it linking algebra} whose diagonal blocks are the open string
vertex algebras $\AA^\str_{a_ia_i}$ for those boundary conditions
$a_i$ related by the action of the
orbifold group $\TT_\fix$, and whose off-diagonal blocks
are the Morita equivalence bimodules $\MM_{a_ia_j}$
mapping $\AA_{a_ia_i}^\str$ to $\AA_{a_ja_j}^\str$
(\ie\ the vertex operators creating strings that
have boundary conditions $a_i$, $a_j$ at either end).
Note that we have a product $\MM_{ab}\otimes_{\AA_{bb}}\MM_{bc} 
\to \MM_{ac}$ so that \Balg\ is indeed an algebra. 
A very similar construction is used in the theory 
of Morita equivalence of $C^*$ algebras
\bgr\raeburn. 

Naively one might expect that $\TT_\fix$
embeds in the linking algebra \Balg\ as a group
of isomorphisms permuting the algebras $\AA_{a_ia_i}$,
however this is generically not the case.
Morita equivalence is {\it not} an isomorphism of algebras,
it is an equivalence relation, which is weaker.
At generic points in Narain moduli space, the linking
algebra has no automorphisms beyond those of
its component algebras along the block diagonal.
An instructive example is the action of the T-duality
$E\rightarrow 1/E$.  The open string algebras depend
parametrically on the Narain data $E$ as well as
the boundary conditions $a_i$; T-duality
acts for example to relate the open string algebra
with Neumann boundary conditions in background $E$
to the open string algebra with Dirichlet boundary
conditions in background $1/E$:
$\AA^\str_{NN}(E)\cong A^\str_{DD}(1/E)$.
This isomorphism relates string algebras with different
boundary conditions at different points in
Narain moduli space.
It is not an automorphism of the linking algebra,
which packages together open string algebras for
different boundary conditions at fixed $E$,
except at the fixed point $E=1/E$.  At such special
points T-duality acts as an automorphism of
the linking algebra $\BB(E_\fix)$.

Thus the linking algebra carries a natural action of $\TT_\fix$,
as an extra group of automorphisms, precisely at
the enhanced symmetry points of Narain moduli space.   
This leads us to conjecture that the desired orbifold
algebra is a direct summand of 
$(\BB(E_\fix)\sdtimes \TT_\fix)^{\TT_\fix}$. 
The regular representations
given by the collections of vertex operators with boundary 
conditions $a\in\{a_i\}$ split into isotypical components
according to the irreducible representations of $\TT_\fix$. 
Presumably, there are other fractional brane projectors,
for instance of the type \anotherproj;
the projection operators should span the lattice
of K-theory charges.
Note that resolving the representations of \Balg\
into isotypical components justifies the formal sums
of $Dp$-branes of different $p$ required to make fractional
branes (particular examples have been studied in
\asymorbrefs).

We leave the nontrivial problem of the 
classification of projectors and the formulation 
of the associated
K-theory of general asymmetric orbifolds to future work.
There is, however, a simple example where we can carry
out the discussion at the level of the zero mode algebra
(in its noncommutative description as $\AA_\Theta$)
in the decoupling limit -- namely,
orbifolds by asymmetric shifts.
In order to formulate it, let us consider first the following
construction. Consider a noncommutative
torus algebra $\widetilde{\AA}_{\widetilde{\Theta}}$
graded by a lattice $\wl\subset \IR^d$.
Let $\dlat\subset \wl$ be a sublattice
with $\wl/\dlat \cong \IZ_N$, so that
\eqn\cosets{
\wl = \dlat \oplus (\dlat+ w) \oplus (\dlat+2w) \oplus \cdots
\oplus
(\dlat+ (N-1)w)
}
where $w\in \wl$ with $Nw\in \dlat$. Let $\AA_{\Theta}$ be the
subalgebra
generated by $U_p$, with $p\in \dlat$. Note that if
$\tilde e_i$ is a basis for $\wl$ then a basis $e_i$ for
$\dlat$ will be given by $e_i = S_i^{~~j} \tilde e_j$ where
$S$ is a matrix of integers with nonzero determinant. Then
$\Theta = S^t \widetilde{\Theta} S$. 
How can we construct $\widetilde{\AA}_{\widetilde{\Theta}}$  from
$\AA_{\Theta}$?
We claim that $\widetilde{\AA}_{\widetilde{\Theta}}$ is the $\IZ_N$
invariant
subalgebra of a linking algebra for $\AA_{\Theta}$. To see this note
that
since $U_w$ normalizes $\AA_{\Theta}$
\eqn\bimodi{
\MM_{mn} := U_{mw} \AA_{\Theta} U_{-nw}
}
is a bimodule for $\AA_{\Theta}$, where $m,n\in \IZ_N$. Indeed, we
could also write it as $\MM_{mn} \cong \MM_{(m-n)\mod N}$ where
\eqn\bimodii{
\MM_{k} := Span\{ U_p| p\in kw + \dlat\}
}
Moreover,  as is evident from \bimodi,
there is a multiplication $\MM_{mn}\otimes \MM_{nm'} \to \MM_{mm'}$.
Therefore, we may use the bimodules \bimodi\ to form the
linking algebra \Balg\  above.
Now, the group $\IZ_N$ acts as a group of automorphisms
on the linking algebra by taking
$B\rightarrow \GG B\GG^{-1}$ where $\GG$ is the element
\eqn\asymshift{
  \GG =\pmatrix{ 0&0&\cdots&\cdots &\gamma\cr
                \gamma&0& & &\vdots \cr
                0&\gamma&\ddots& &\vdots\cr
                \vdots& &\ddots&0&0\cr
                0&\cdots &\cdots&\gamma&0}
        \quad,\qquad
        \gamma=U_w\ .
}
Now we have
\eqn\claim{
\widetilde{\AA}_{\widetilde{\Theta}} =  \BB^{\IZ_N}\ .
}
%
%

The lattice of momenta $\wl$ is precisely what results
from the orbifold by an asymmetric shift by a fractional winding
$w$, such that $Nw$ is an allowed closed string winding vector
in the Narain lattice (the latter is isomorphic to the unique
self-dual Lorentzian lattice $II^{d,d}$).
Of course, in this example we could have used the 
Morita equivalence to the T-dual algebra $\AA_{1/\Theta}$,
and performed a symmetric shift \symorbact\ there;
however, in the general orbifold group including both
asymmetric twists and shifts, we will need to consider
algebras $\Balg$ containing T-dual algebras and this
procedure will not be available.

It would be interesting to consider the
most general asymmetric shift $v$, $Nv\in II^{d,d}$;
however, this would take us into a lengthy detour into
the specifics of shift vectors, level matching 
constraints, and the like, and so we will not pursue it here.
A few useful remarks about orbifolds by
shifts are collected in Appendix B.


\newsec{Conclusion: Some future directions} 

D-branes are sources of RR fields and hence carry RR charge. 
The RR charges are neatly summarized by the  Chern-Simons 
coupling in the D-brane worldvolume Lagrangian. It is natural 
to ask whether such couplings can be formulated in the context 
of noncommutative geometry. We plan to address this question in 
\paperii. 

In this paper we have examined rather simple orbifolds and 
crossed-product algebras. Given the examples which tend to 
be studied in the $C^*$-algebra literature it is natural to 
ask whether orbifolds by other infinite discrete groups 
(for example, non-amenable groups) or by ergodic actions 
of real Lie groups might provide interesting examples of 
string backgrounds. A key requirement in formulating orbifolds 
in string theory is that the orbifold group must be a symmetry 
of the dynamics, so that we can gauge it. 
For this reason, it is unlikely that one 
can make a sensible string background based on foliations of 
tori. Nevertheless, there are some backgrounds and limits of 
string theory where the Hamiltonian is effectively zero and 
where one 
might consider more nontrivial crossed-product algebras. 
One particularly interesting example might be formulating 
string theory on quotients of $T^*SL(2,\RR)$ by infinite 
discrete groups of hyperbolic isometries.  
Similar quotients were considered as cosmological models
in \hormar.
If such backgrounds
make sense then the viewpoint of this paper should prove 
useful for the formulation of the corresponding D-branes. 
Such orbifolds will not preserve any supersymmetry, and
generically will be expected to have tachyons;
nevertheless, the classical string theory is well-defined
and might be of some interest.


\bigskip\medskip\noindent
{\bf Acknowledgements:}
We wish to thank
M. Douglas,
R. Douglas,
A. Hashimoto,
N. Itzhaki,
F. Larsen,
H. Liu, 
P. Kraus,
B. Pioline,
and
A. Schwarz
for helpful conversations.
We especially thank J. Harvey for collaboration in the 
initial stages of the project.
This work was supported by 
DOE grants 
DE-FG02-90ER-40560 and
DE-FG02-96ER-40949.


\appendix{A}{Morita equivalence bimodules on higher-dimensional tori}

Here we summarize results of Rieffel \rieffel\ on the construction
of modules for the noncommutative torus algebra
$\AA_\Theta$ \gentoralg\ on tori of arbitrary dimension.
Consider somewhat more generally any locally compact abelian group
$M$, its dual group $\hat M$, and let $G=M\times\hat M$.
On $G$ there is a canonical cocycle
\eqn\canco{
  \beta\bigl((s,t),(v,w)\bigr)=\exp[2\pi i\vev{s,w}]
	\quad,\qquad s,v\in M\ ,\quad t,w\in\hat M\ ,
}
and a projective representation of $G$
on the space of Schwarz functions $\SS(M)$ via
\eqn\ltworep{
  U_{(s,t)}^{ }f(v)=e^{2\pi i\vev{v,t}}\;f(v+s)\ .
}
{}From this one easily deduces
\eqn\heisalg{ \eqalign{
  U_{ x}U_{ y}&=\beta( x, y)U_{ x+ y}\cr
  U_{ x}U_{ y}&=\beta( x, y)\bar\beta( y, x) \; U_{ y}U_{ x}
	\equiv \rho(x,y) \; U_{ y}U_{ x}
}}                            
for $x=(s,t),y=(v,w)\in G$.

Now take $M=\IR^p\times\Z^q\times \Z_r$
(so $\hat M=\hat\IR^p\times\T^q\times\hat \Z_r$),
and $2p+q=d$; and represent it as in \ltworep\ on $\SS(M)$.  
Suppose $\dlat\subset G$ is a cocompact group, with 
$\dlat\cong \IZ^d$, and such
that the elements $U_{(s,t)}$ with $(s,t)\in \dlat$ generate $\AA_{\Theta}$.
Then, since $\dlat$ is cocompact, the orthogonal complement determined by 
\eqn\dperp{
  \dlat^\perp=\{y\in G:\rho(x,y)=1\quad\forall\  x\in \dlat\}
}
defines an algebra $\BB=\AA_{\Theta'}$
Morita equivalent to $\AA=\AA_\Theta$.  
In the 2d example of section 4, 
we had $p=1$, $q=r=0$;
thus $G=\IR\times\hat\IR$, with cocycle
\eqn\rhotwod{
  \rho_{2d}^{ }((s,t),(v,w))=\exp[2\pi i(sw-tv)]\ ;
}
the lattices $\dlat$ and $\dlat^\perp$ are 
$\dlat=\{(s,t)=(j\theta,k),~j,k\in\Z\}$ and 
$\dlat^\perp=\{(v,w)=(m,n/\theta),~m,n\in\Z\}$.
Indeed the algebra $\AA_{\theta'}$ defined by $\dlat^\perp$ is 
Morita equivalent to $\AA_\theta$ by $\theta\rightarrow -1/\theta$.%
\foot{The representation of the algebra \bprod\ is the conjugate
or opposite algebra of \ltworep\ 
because it acts oppositely (on the left vs. the right).}
The construction agrees with the analysis of \sw\
which showed that the zero modes of 2-2 strings live in
a Hilbert space of functions on $\T^2$ 
(the Fourier dual to $\Z^2$ used above for $p=0$, $q=2$), 
whereas the zero modes of 2-0 strings live in a Hilbert space
of functions on $\IR$.  Physically, the $NN$ coordinates
describe quantized momenta, the $DD$ coordinates have quantized winding,
and hence their zero mode momenta or winding
lives on the appropriate lattice. 
On the other hand, the $DN$ strings' zero modes
describe a rigid rotator tethered on one end, whose
free end lives on the covering space of the torus.
The wavefunction for the free end will thus 
depend on only half of the coordinates.
Thus the condition $2p+q=d$.

Functions $f,g\in \SS(M)$ yield elements of the algebra $\AA=\AA_\Theta$
just as in the two-dimensional case \aprod\ via
\eqn\arep{
  \vev{f,g}_\AA=\sum_{(s,t)\in \dlat}\Bigl(\int_M dv 
	\overline{f(v+s)}g(v)e^{2\pi i\vev{v,t}}\Bigr)
	\;U_{(s,t)}^{ }\ ,
}
and similarly for $\BB=\AA_{\Theta'}$; the Morita equivalence
relation
\eqn\moreq{
  \vev{f,g}_\BB^{ }h=f\vev{g,h}_\AA^{ }
} 
for $f,g,h\in\SS(M)$, follows by a Poisson summation formula
relating sums over $\dlat$ and $\dlat^\perp$ \rieffel.

Now one may choose a {\it finite set} \rieffel\ of 
elements $f_1,...,f_n\in\SS(M)$ such that
\eqn\bident{
  \sum_i\vev{f_i,f_i}_\BB^{ }=\One_\BB\ ;
}
%
%
then the $\AA$-valued inner products
\eqn\aproj{
  \P_{ij}=\vev{f_i,f_j}_\AA^{ }
}
are the matrix elements of a projector 
$\P\in{\rm Mat}_n(\AA)$, and \rieffel\ shows
that all $\AA$ modules are obtained in this way.
In this construction, one finds that $2p$ is the rank
of the highest nonzero Chern class of the module; the role
of the finite group $\Z_r\times\hat \Z_r\subset G$ is to allow
for twisted boundary conditions,
as in the discussion of boundary conditions for 2-($n$,$m$) strings in
section 6.3 of \sw\ 
(see also \refs{\connescrasp,\schmo,\mz,\hv,\koscorb}).


\appendix{B}{Comments on shifts}

One way to write the Narain lattice is as follows.
Take the standard Euclidean metric 
on $\IR^d \oplus \IR^d$ of signature $(1,-1)$. Then
\eqn\narlat{\eqalign{
  e^a & = {1\over \sqrt{2}} (e^a_{~~\nu} \; ; \;  e^a_{~~\nu})\cr
  f_a & = {1\over \sqrt{2}} (e_a^{~~\nu}E_{\nu\mu} \; ; \; 
         -e_a^{~~\nu}E_{\nu\mu}^{\,t}) 
}}
with $a,\mu=1,\dots, d$, 
span a lattice isomorphic to $II^{d,d}$. 
(Here take $g_{\mu\nu} = \delta_{\mu\nu}$, and $E=g+B$.)  
Put differently,
\eqn\Edefn{
  {\bf E} :=
        {1\over \sqrt{2}} \pmatrix{ e^a_{~~\nu} &  e^a_{~~\nu} \cr
        e_a^{~~\nu}E_{\nu\mu} &  e_a^{~~\nu}E_{\nu\mu}^{t} \cr}
}
satisfies
\eqn\Eprop{
  {\bf E} \pmatrix{1& 0 \cr 0 & -1 \cr} {\bf E}^{t} = 
        \pmatrix{0 & 1 \cr 1 & 0 \cr}
}
The lattice metric is $g^{ab}=e^a_{~~\mu} e^b_{~~\mu}$,  
where  $e_a^{~~\mu} e^b_{~~\mu}= \delta_a^{~b}$.

Let $p=m_a e^a + w^a f_a$, $m_a, w^a\in\IZ$
be the generic lattice vector in the
lattice spanned by $e^a, f_a$. 
Suppose we have a shift vector so that $Nv \in II^{d,d}$ is primitive. 
 Then, provided the level matching condition $N v^2\in 2\IZ$ 
is satisfied,   the lattice defined by taking the projection
$p\cdot v\in \IZ$ together with the union of cosets $(p+\ell v)$ 
(projected to the vectors with  $(p + \ell v)^2\in 2\IZ$)
defines another even unimodular lattice.  This lattice will be
spanned by $\bar e_a, \bar f^a$ related to
the original basis vectors by
\eqn\Nspan{
  \pmatrix{ \bar e^a \cr \bar f_a} = S\pmatrix{  e^a \cr  f_a}
}
where $S$ is a rational matrix such that
\eqn\Sprop{
   S \pmatrix{0 & 1 \cr 1 &  0 \cr} S^{t} = \pmatrix{0 & 1 \cr 1 &  0 \cr}
}
One way  to prove this is 
to use the Narain-Siegel theta function.
The shift has thus produced an orbifold at another point
in Narain moduli space specified by ${\bf E}' = S {\bf E}$.

To make the above construction more explicit we need to 
say what $S$ is.  We can be slightly
more explicit if we consider an orbifold with a shift of the form
$v = { 1\over N} v^a f_a$, with $v^a\in \IZ$.
The condition $p\cdot v= 0~\mod~1$, becomes $m_a v^a = 0~\mod~N$.
This  condition defines an integral sublattice 
of the $d$-dimensional lattice
$m_a e^a$. This new lattice will be spanned by vectors
$\bar e^a = s^{a}_{~~b} e^b$ where the matrix $s$ is a matrix of
integers. Let the   d-dimensional lattice spanned by
the $\bar e^a$ be denoted by $L$.

Now let $F$ be the lattice spanned by $f_a$.  
Then $F, F+v, F+2v, \dots$ spans a Euclidean lattice.  
We claim this is the dual lattice $L^*$. 
One way to prove this is the following. 
For the purposes of this argument we introduce an 
auxiliary metric on the span of $f_a$ such that 
$\langle\langle f_a, f_b \rangle \rangle = \delta_{ab}$
and form the theta function of the lattice. Similarly we 
introduce such an auxiliary metric on $L$, that is on 
the span of $e^a$,  $\langle\langle e^a,e^b \rangle \rangle = \delta^{ab}$. 
Then from the definition of $L$ we have  
\eqn\thetalat{
  \Theta_L = \sum_{m_a\in \IZ} e^{i \pi \tau m_a^2} \biggl({1\over
        N}\sum_{\ell=0}^{N-1}
        e^{2\pi i m_a v^a \ell/N}\biggr)
}
On the one hand, $\Theta_L(-1/\tau) = \sqrt{|L/L^*|}\Theta_{L^*}(\tau)$;
on the other hand, doing the modular transformation term-by-term gives
the description in terms of $F$ and its translates by $v$.
The lattice $F \oplus (F+v) \oplus (F+2v) \oplus \cdots$ has a basis
$\bar f_a = (s^{tr,-1})_a^{~b} f_b$.

Thus, the lattice
after the shift is spanned by vectors $\bar e^a, \bar f_a$ which
are related to the original vectors $e^a, f_a$ by
\eqn\vecrel{
  \pmatrix{ \bar e^a \cr \bar f_a} = 
        \pmatrix{s & 0 \cr  0 & (s^t)^{-1} \cr} 
        \pmatrix{  e^a \cr  f_a}
}
Note that the matrix is a rational matrix in $O(d,d)$.
Now, the new point in Narain moduli space is
\eqn\newpoint{
{\bf E}' = \pmatrix{s & 0 \cr  0 & (s^t)^{-1} \cr} {\bf E}
}
This is just an $GL(d,\IQ)$ transformation of the torus, holding
$B_{\mu\nu}$ in the covering space coordinates fixed.
Similar remarks apply to shift vectors which are
of the form $v = { 1\over N} v_a e^a$.
 
For example, consider $v= {1\over N} (a e + b f)$ in the 1d case.
Let us suppose that $ab$ is nonzero. 
By level matching $ab$ divides $N$.
Allowed momenta in the untwisted sector satisfy $p\cdot v\in\IZ$,
for $p=ke+wf$, $k,w\in\IZ$; in other words,
$k\in\frac{N}{(N,b)}\IZ$, $w\in\frac{N}{(N,a)}\IZ$.
The twisted sectors add all $k\in\frac{(N,a)}{(N,b)}\IZ$
and $w\in\frac{(N,b)}{(N,a)}\IZ$, so that the net effect
is that the radius of the circle has been rescaled by
$s=\frac{(N,b)}{(N,a)}$.


\listrefs
\end